\documentstyle[epsfig,times]{mn}

 at 10truept

\title[The shear power spectrum from the COMBO-17 survey]
  {The shear power spectrum from the COMBO-17 survey}
\author[M.L.~Brown et al.]
  {M.L.~Brown$^1$\thanks{mlb@roe.ac.uk; ant@roe.ac.uk; djb@roe.ac.uk},
  A.N.~Taylor$^{1\star}$, 
  D.J.~Bacon$^{1\star}$, 
  M.E.~Gray$^1$, S.~Dye$^2$,
  K.~Meisenheimer$^3$, \newauthor
  \& C.~Wolf$^4$\\
  $^1$Institute for Astronomy, University of Edinburgh,
      Blackford Hill, Edinburgh EH9 3HJ \\
    $^2$Astrophysics Group, Blackett Laboratory,
      Imperial College, Prince Consort Road, London SW7 2BW\\
  $^3$Max-Planck-Institut f\"ur Astronomie, K\"onigstuhl 17,
       D-69117, Heidelberg, Germany \\
  $^4$Department of Physics,
      University of Oxford, Keble Road, Oxford OX1 3RH \\
\\}

\date{Accepted 2002 November 29. Received 2002 November 12; in original form 2002 October 9}

\pagerange{100--118} \pubyear{2003} \volume{341}

\def\LaTeX{L\kern-.36em\raise.3ex\hbox{a}\kern-.15em
    T\kern-.1667em\lower.7ex\hbox{E}\kern-.125emX}

\def\bib{\parskip=0pt\par\noindent\hangindent\parindent
    \parskip =2ex plus .5ex minus .1ex}

\newcommand{\be}{\begin{equation}}
\newcommand{\ee}{\end{equation}}
\newcommand{\ba}{\begin{eqnarray}}
\newcommand{\ea}{\end{eqnarray}}

\newcommand{\nn}{\nonumber \\}
\newcommand{\nnb}{\begin{displaymath}}
\newcommand{\nne}{\end{displaymath}}
\newcommand{\de}{\partial}

\newcommand{\hMpc}{\,h^{-1}{\rm Mpc}}
\newcommand{\Mpc}{\,h^{-1}{\rm Mpc}}
\newcommand{\hkpc}{\, h^{-1}{\rm kpc}}

\newcommand{\r}{\mbox{\boldmath $r$}}
\newcommand{\rhat}{\hat{\r}}
\newcommand{\lb}{\mbox{\boldmath $\ell$}}
\newcommand{\lbh}{\mbox{\boldmath $\hat{\ell}$}}

\newcommand{\thetab}{\mbox{\boldmath $\theta$}}
\newcommand{\rgl}{\rangle}
\newcommand{\lgl}{\langle}

\newcommand{\rh}{\hat{r}}

\newcommand{\C}{\mbox{\boldmath $C$}}
\newcommand{\N}{\mbox{\boldmath $N$}}
\newcommand{\I}{\mbox{\boldmath $I$}}

\newcommand{\db}{\mbox{\boldmath $d$}}

\begin{document}

\maketitle

\begin{abstract}
We perform a statistical weak lensing  or ``cosmic shear''
analysis of the COMBO-17 survey -- a unique dataset with shear 
quality R-band imaging and accurate photometric redshift 
estimates ($\sigma_z=0.05$) for $\sim90\%$ of  galaxies to 
$m_{\rm{R}} \le 24.0$. We undertake a full maximum
likelihood analysis to measure directly from the data the weak
lensing power spectra, $C^{\kappa\kappa}_\ell$,
$C^{\beta\beta}_\ell$ and $C^{\kappa\beta}_\ell$ in 5
band powers from $\ell=400$ to $\ell=10^4$, where
$\kappa$ is the usual lens convergence and $\beta$ is an
odd-parity `curl' component of the shear signal. We find a strong
measurement of the convergence power over five fields. The
non-gravitational $\beta$-field has a much lower significance,
indicating our data is free of major systematics, while the
cross-correlation of $\kappa$ and $\beta$ is consistent with zero.
We have also calculated the shear correlation functions and variance 
over a range of scales between $0.5$ and $20$ arcmin. Our measurements
of minimal star-galaxy correlations and cross-correlations between 
galaxy components provide further evidence that any systematics are 
negligible.

In addition, we have used our results to measure cosmological 
parameters, constraining the normalisation of the matter power 
spectrum to be $\sigma_8=(0.72 \pm 0.09)(\Omega_m/0.3)^{-0.49} $, 
where the errors quoted are 1-$\sigma$ due to the intrinsic dispersion 
in galaxy ellipticities, cosmic and sampling variance. We have 
significantly reduced the usual additional uncertainty in the median 
redshift ($z_m$) of the source galaxies by estimating $z_m$ directly 
from our data using accurate photometric redshift information from the 
COMBO-17 multi-band wide-field survey. To demonstrate the power of 
accurate redshift information, we have also measured parameters from a 
shear analysis of only those galaxies for which accurate redshift estimates 
are available. In this case, we have eliminated the uncertainty in the redshift 
distribution of sources and we show that the uncertainty in the resulting 
parameter constraints are reduced by more than a factor of 2 compared to the
typical uncertainties found in cosmic shear surveys to date. Finally, we 
combine our parameter measurements with constraints from the 2dF Galaxy 
Redshift Survey and with those from the CMB. With these additional 
constraints, we measure the normalisation of the matter power spectrum 
to be $\sigma_8=0.73^{+0.06}_{-0.03}$ and the matter density of the 
Universe to be $\Omega_m=0.27^{+0.02}_{-0.01}$. 
\end{abstract}


\label{firstpage}
\section{Introduction}

There continues to be great interest and significant progress in
measuring the weak lensing signal arising from large-scale
structure. This phenomenon, observed as the weak coherent distortion
of background galaxies due to light ray deflection by intervening
matter, offers us a direct probe of the mass distribution in the
Universe. We can consequently measure cosmologically important
quantities (see e.g.. Bernardeau et al. 1997; Jain \& Seljak 1997;
Kamionkowski et al. 1998; Kaiser 1998; Hu \& Tegmark 1999) such as the
bias (see e.g. Hoekstra et al. 2001) and the normalisation of the
matter power spectrum (e.g. Bacon et al. 2002; Hoekstra et al. 2002;
Refregier, Rhodes \& Groth 2002; Van Waerbeke et al. 2002).

Of particular interest is the prospect of reconstructing the power
spectrum of mass fluctuations from the weak lensing signal. Pen et al.
(2002) have used an estimator for the weak lensing shear power spectrum, 
obtainable from correlation function measurements, to measure the shear 
power spectrum from the DESCART survey, while Schneider et al. (2002) have 
further developed estimators for this purpose. Meanwhile, Hu \& White 
(2001) have demonstrated the utility of a maximum likelihood approach to
reconstructing the shear power spectrum. Here we will apply a full maximum
likelihood analysis to a cosmic shear survey for the first time.

In this paper, we describe the results of a weak shear analysis of the
COMBO-17 dataset, acquired with the La Silla 2.2m telescope in Chile
(Wolf et al. 2001). This survey includes 1.25 square degrees of deep 
${\rm R}$-band observations from which we draw our sample of background
galaxies for shear measurements. In addition, the survey has yielded 
photometric redshifts for approximately 40\% of the objects in our 
selected sample of galaxies, which we use to improve our understanding 
of the shear signal. We apply a maximum likelihood reconstruction to our 
dataset to obtain the shear power spectrum, and we fit cosmological models 
to measurements of both the shear power spectrum and the correlation 
functions to measure joint constraints on the matter density $\Omega_m$ 
and normalisation of the matter power spectrum $\sigma_8$. We obtain 
further constraints on these parameters by combining our measurements 
with constraints from both the 2dF Galaxy Redshift Survey (2dFGRS, 
Percival et al. 2002) and from CMB data (e.g. Lewis \& Bridle, 2002).

Cosmic shear studies require careful data reduction followed by
assessment and removal of systematic effects, such as shear induced by
a telescope, anisotropic point spread function (PSF), and circularization of 
galaxy shapes. We demonstrate in this paper that our shear catalogues have 
almost negligible systematic errors after applying appropriate corrections, and
we carefully audit the remaining sources of error (from e.g. sample size,
redshift uncertainty and shot noise) in order to interpret our results.

The paper is organized in the following fashion. In Section 2 we
summarize the necessary formalism for our weak lensing analysis,
including definitions of shear and the shear power spectrum. In
Section 3 we describe the COMBO-17 survey. We then discuss the
procedures used for data reduction, and describe our method for
obtaining a shear catalogue fully corrected for systematic
effects. Section 4 examines the individual COMBO-17 fields in detail,
and explains our approach to including redshift information for these
fields. We measure shear correlation functions and cell variance in
Section 5, checking for residual systematics and comparing with the
findings of other groups. Section 6 contains our likelihood analysis,
resulting in shear power spectra measured for single fields and for the 
survey as a whole. In Section 7 we use our correlation functions and power
spectra measurements to estimate cosmological parameters, taking into 
account the errors  from sample size and redshift uncertainties. We then
combine our constraints on these parameters with constraints obtained 
from the 2dFGRS and the CMB. In Section 8 we discuss these results and 
summarize our conclusions.

\section{Weak lensing quantities}

In the last few years weak gravitational lensing has emerged as
the most direct method for measuring the distribution of matter,
regardless of its nature, in the Local Universe (Mellier 1999, 
Bartelmann \& Schneider 2001). This is largely due to its basis 
in well understood physics: weak lensing is essentially a scattering 
experiment of photons off the gravitational field generated by 
cosmological structure. Here, we describe the various weak lensing 
fields and their statistical properties. 

\subsection{Weak lensing fields}

Weak gravitational lensing induces a distortion into the images of
distant source galaxies. This distortion can be parameterised by
measuring the ellipticities,
 \be
    e_{ij} =\left( \begin{array}{cc} e_1 & e_2 \\
                                 e_2 & -e_1 \end{array}
                                                        \right)
 \ee
of each galaxy from its quadrupole moments at a given isophotal 
threshold (Kaiser, Squires and Broadhurst, 1995) or by examining 
the distortion to a set of orthogonal modes describing the galaxy 
shape (Refregier \& Bacon 2003; Bernstein \& Jarvis, 2002). 
In the first instance -- which is the approach we adopt in the 
following analysis -- the effect of lensing is to induce an additional 
ellipticity on the galaxy image;
 \be
    e'_{ij} = e_{ij} + 2 \gamma_{ij},
    \label{ellip}
 \ee
 where
 \be
    \gamma_{ij} =\left( \begin{array}{cc} \gamma_1 & \gamma_2 \\
                                 \gamma_2 & -\gamma_1 \end{array}
                                                        \right)
 \label{shearmatrix}
 \ee
 is the trace-free lensing shear matrix. The components $\gamma_1$ 
 and $\gamma_2$ of the shear matrix represent the two orthogonal 
 modes of the distortion. The shear matrix is recoverable 
 since in the absence of intrinsic alignments of galaxies, the 
 galaxy ellipticities average to zero, $\lgl e \rgl =0$.

 Since gravity is a potential theory in the weak-field regime, the
 shear field can be related to a lensing potential
  \be
        \gamma_{ij} = \left( \de_i \de_j - \frac{1}{2}
        \delta^K_{ij}\de^2 \right) \phi,
        \label{shear}
  \ee
  where $\de_i  \equiv r (\delta_{ij}- \rh_i \rh_j) \nabla_i $
  is a dimensionless, transverse differential operator, and $\de^2
\equiv \de_i \de^i$ is the transverse Laplacian. The indices
$(i,j)$ each take the values $(1,2)$, and we have assumed a flat 
sky. On the scales currently of interest (i.e. from $\sim100\hkpc$ to 
$100\hMpc$), this is an excellent approximation.

The lensing potential is also observable via the lens convergence
field
 \be
    \kappa = \frac{1}{2} \de^2 \phi,
 \ee
 which can be estimated from the weak magnification of sources,
 \be
    \mu = |(1-\kappa)^2 - \gamma^2|^{-1} \approx 1+ 2 \kappa.
 \ee
 This can be measured from either the change in galaxy number
 density (Broadhurst, Taylor \& Peacock 1995; Taylor et al. 1998)
 or from the change in size of sources (Bartelmann \& Narayan, 1995). 
 Although this has a lower signal-to-noise than the shear distribution, 
 it is an independent estimator of the lensing potential.

 The convergence field is related to the shear field by the
 differential relation, first used by Kaiser \& Squires (1993);
  \be
        \kappa = \de^{-2}\de_i \de_j \gamma_{ij},
        \label{kappa}
  \ee
 where $\de^{-2}$ is the inverse 2-D Laplacian operator defined by
 \be
        \de^{-2} \equiv \frac{1}{2\pi} \int \! d^2 \!\rhat' \, \ln |\rhat - \rhat'|.
 \ee

A useful quantity for tracing noise and systematics in
gravitational lensing is the divergence-free field, $\beta$,
defined by
 \be
	\beta = \de^{-2} \varepsilon^n_{i} \de_{j} \de_n \gamma_{ij}.
        \label{beta}
 \ee
 where $\varepsilon^n_{i}$ is the Levi-Civita symbol in two 
 dimensions,
 \be
	\varepsilon^n_{i} =\left( \begin{array}{cc} 0 & -1 \\
                                 1 & 0 \end{array}
                                                        \right).
 \ee  

 If $\gamma_{ij}$ is generated purely by the lensing potential,
 this quantity vanishes. But if there are non-gravitational sources,
 due to noise, systematics or intrinsic alignments, $\beta$ will be
 non-zero. In addition $\beta$ terms can arise from finite fields,
 due to mode-mixing of the $\kappa$ and $\beta$ fields (Bunn, 2002).

In a spatially flat Universe, with comoving distance $r$ we can
relate the lensing potential and the gravitational potential by
(e.g. Kaiser 1998; Hu 2000)
\begin{equation}
\phi(\r) = 2 \int_0^{r}\!\! dr' \, \left(\frac{r-r'}{r r'}\right)
\Phi(\r'). \label{pot}
\end{equation}
This equation assumes the Born approximation, in which the path of
integration is unperturbed by the lens.

Although the lensing potential depends on the distance to the
source galaxy, this dependence is usually lost by averaging all
lensing quantities over the source distribution. However with
redshift information the full 3-D character of lensing can be
usefully recovered (Taylor 2001, 
Bacon \& Taylor 2002, Hu \& Keeton 2002).

Finally the Newtonian potential, $\Phi$, can be related to
perturbations in the matter density field, $\delta = \delta
\rho_m/\rho_m$, by Poisson's equation;
\begin{equation}
\nabla^2 \Phi = 4 \pi G \rho_m \delta a^2 = \frac{3}{2}
\frac{\Omega_m}{ a \lambda_H^2} \delta,
\end{equation}
where we have introduced the cosmological scale factor $a$, the
Hubble length $\lambda_H=c/H_0 \approx 3000 \Mpc$, and the
present-day mass-density parameter $\Omega_m$. Here, $H_0$ is the
Hubble constant and $c$ is the speed of light.

\subsection{Statistical properties}

\subsubsection{Shear covariance matrix}

We may define a shear covariance matrix by
 \be
    C_{ab}(\rhat) = \lgl \gamma_a(0) \gamma_b(\rhat) \rgl,
\label{shear_signal}
 \ee
where the indices $(a,b)$ each take the values (1,2). 
Fourier transforming the shear field,
 \be
    \gamma_{ij} (\lb) = \int \! d^2 \rhat \,
        \gamma_{ij}(\rhat) e^{- i \lb . \rhat },
 \ee
and decomposing it using equations (\ref{kappa}) and (\ref{beta}), 
we may generate the shear power spectra from correlations of $\kappa$ 
and $\beta$:
 \ba
    \lgl \kappa(\lb) \kappa^*(\lb') \rgl &=& (2 \pi)^2 C^{\kappa
    \kappa}_\ell \delta_D(\lb - \lb'), \nn
       \lgl \beta(\lb) \beta^*(\lb') \rgl &=& (2 \pi)^2 C^{\beta
    \beta}_\ell \delta_D(\lb - \lb'), \nn
       \lgl \kappa(\lb) \beta^*(\lb') \rgl &=& (2 \pi)^2 C^{\kappa
    \beta}_\ell \delta_D(\lb - \lb').
 \ea
The parity invariance of weak lensing suggests that
$C^{\beta\beta}_\ell=C^{\kappa\beta}_\ell=0$. However other
effects, such as noise and systematics, as well as intrinsic
galaxy alignments may give rise to a non-zero
$C^{\beta\beta}_\ell$. Hence in our analysis we shall leave its
amplitude to be determined by the data. The cross-correlation of 
$\kappa(\ell)$ and $\beta(\ell)$ is expected to be zero but it will 
also allow a second check on noise and systematics in the
shear field, and we shall treat is as another free function. In
particular finite field and boundary effects can lead to leakage of power
between these three spectra, which we shall attempt to monitor.

The shear power spectrum and the convergence power are related by
  \be
         C^{\gamma \gamma}_\ell = C^{\kappa\kappa}_\ell
  \ee
in the flat-sky approximation. For a spatially flat Universe, 
these are in turn related to the matter power spectrum, 
$P_{\delta}(k,r)$ by the integral relation (see e.g. Bartelmann 
\& Schneider 2001):
  \be
	C^{\kappa \kappa}_\ell = \frac{9}{4} \left(\frac{H_0}{c}\right)^4 
	\Omega_m^2\, \int_0^{r_{\mathrm H}}\! dr \,
	P_\delta \left (\frac{\ell}{r},r\right)
	\left(\frac{\overline{W}(r)}{a(r)}\right)^2 ,
	\label{eq:gampowspec}
  \ee
where $a$ is the expansion factor and $r$ is comoving distance.
$r_{\rm H}$ is the comoving distance to the horizon:
  \be
	r_{\rm H}= c \int \frac{dz}{H(z)},
  \ee  	
where the Hubble parameter is given in terms of the matter density, 
$\Omega_m$, the vacuum energy density, $\Omega_V$ and the spatial
curvature, $\Omega_K$ as
  \be
	H(z)=H_0[(1+z)^3 \Omega_m+(1+z)^2\Omega_K+\Omega_V]^{1/2}.
  \ee
The weighting, $\overline{W}$, is given in terms of the normalised
source distribution, $G(r) dr = p(z) dz$:
 \be
   \overline{W}(r) \equiv \int_r^{r_{\mathrm H}}\! dr'\,G(r')\,
   \frac{r'-r}{r'}.
 \ee

The covariance of the components of the shear field are related to
the power spectra by (Hu \& White 2001)
 \ba
    C_{11}(\rhat) &=& \int \! \frac{d^2 \ell}{(2 \pi)^2} \,
    \big( C^{\kappa\kappa}_\ell \cos^2 2 \varphi_\ell +
            C^{\beta \beta}_\ell \sin^2 2 \varphi_\ell \nn
             & &\hspace{2.cm} -
            C^{\kappa \beta}_\ell \sin 4 \varphi_\ell \big)|W(\ell)|^2
            e^{i \lb.\rhat}, \nn
    C_{22}(\rhat) &=& \int \! \frac{d^2 \ell}{(2 \pi)^2} \,
    \big( C^{\kappa\kappa}_\ell \sin^2 2 \varphi_\ell +
            C^{\beta \beta}_\ell \cos^2 2 \varphi_\ell \nn
             & &\hspace{2.cm} +
            C^{\kappa \beta}_\ell \sin 4 \varphi_\ell \big)|W(\ell)|^2
            e^{i \lb.\rhat}, \nn
    C_{12}(\rhat) &=& \int \! \frac{d^2 \ell}{(2 \pi)^2} \,
    \big( \frac{1}{2}(C^{\kappa\kappa}_\ell -
            C^{\beta \beta}_\ell) \sin 4 \varphi_\ell \nn
             & &\hspace{2.cm}+
            C^{\kappa \beta}_\ell \cos 4 \varphi_\ell \big)|W(\ell)|^2
            e^{i \lb.\rhat},
 \label{shearcovar}
 \ea
 where $\cos \varphi_\ell = \lbh . \lbh_x$ and $\lbh_x$ is a
 fiducial wavenumber projected along the x-axis. We have included here
 a smoothing or pixelisation window
 function;
 \be
    W(\lb) = j_0(\ell_x \theta_{\rm pix}/2)j_0(\ell_y \theta_{\rm pix}/2),
 \ee
 where $j_0=\sin(x)/x$ is the zeroth order spherical Bessel
 function and $\theta_{\rm pix}$ is the smoothing/pixel scale.

\subsubsection{The rotated shear correlation function}
Another important two-point statistical measure of the weak lensing signal are
the correlation functions, defined by 
 \ba
	C_1(\theta) &=& \lgl \gamma_1^r(\rhat)\gamma_1^r(\rhat+\thetab) \rgl, \nn 
        C_2(\theta) &=& \lgl \gamma_2^r(\rhat)\gamma_2^r(\rhat+\thetab) \rgl, \nn
        C_3(\theta) &=& \lgl \gamma_1^r(\rhat)\gamma_2^r(\rhat+\thetab) \rgl,  
 \label{corrfns}
 \ea
where the angled brackets denote the average over all galaxy pairs separated 
by an angle $\theta$. The superscript, $r$ denotes rotated shear components
which are equivalent to $\gamma_1$ and $\gamma_2$ in a rotated coordinate 
frame,  defined by the line joining the centroids of the two galaxies in 
question. In terms of the elements of the shear matrix 
(equation (\ref{shearmatrix})), these rotated shear components are    
 \ba
        \gamma_1^r &=&  \gamma_1\cos(2\phi)+\gamma_2\sin(2\phi) \nn
        \gamma_2^r &=&  -\gamma_1\sin(2\phi)+\gamma_2\cos(2\phi),
 \ea
where $\phi$ is the angle between the original and rotated 
coordinate frames. Once again, the parity invariance of weak lensing 
predicts that the cross-correlation function $C_3(\theta)$ should 
be zero. A non-zero $C_3(\theta)$ is, therefore, an indication of residual 
systematic effects present in the data. The correlation functions, $C_1(\theta)$ 
\& $C_2(\theta)$ are related to the underlying convergence power spectrum via
 \be
    C_{i} (\theta) =
        \int_0^{\infty} \! \frac{d \ell\, \ell}{4 \pi} C^{\kappa \kappa}_{\ell} 
        \left[ J_0(\ell\theta)+(-1)^{i+1} J_4(\ell\theta) \right],
 \label{corr2cl}
 \ee
where $J_n(x)$ are Bessel functions.

\subsubsection{Shear variance in cells}

Finally, we can define the variance, $\sigma^2_\gamma(\theta)$ 
of the shear field, measured in circular cells of radius $\theta$.
In terms of the convergence power spectrum, this variance is 
 \be
        \sigma^2_\gamma(\theta) = \frac{1}{2\pi}\int_0^{\infty}\! d \ell\, \ell
        C^{\kappa\kappa}_{\ell}
        \left(\frac{2 J_1(\ell\theta)}{\ell\theta}\right)^2.
 \label{var2cl}
 \ee
In the following analysis, we have chosen to measure the shear variance in 
square cells of side length, $\theta$. However, the corresponding variance 
in circular apertures can be recovered, to a good approximation, by scaling 
our measurements by $1/\sqrt{\pi}$ (e.g. Bacon, Refregier \& Ellis 2000). 

Having considered the shear field and its statistical properties we
now turn to our data set. In Sections 3 \& 4 we describe the COMBO-17 
dataset and our method used in producing the high quality deep {\rm R}-band 
images from which we measure the shear field. In Sections 5 \& 6  we shall 
use this data to determine the statistical properties of the observed shear field.

\section{Observations and Data Reduction}
\subsection{The COMBO-17 survey}
The observations analysed in this paper have been undertaken as part of the COMBO-17 survey
(Wolf et al. 2001). This survey is being carried out with the Wide-Field Imager (WFI) at the MPG/ESO
2.2m telescope on La Silla, Chile. The survey currently consists of five $0.5\degr \times 0.5\degr$ 
fields totaling 1.25 square degrees with observations taken in five broad-band filters ($UBVRI$) and 12
narrow-band filters ranging from $420$ to $914$ nm. The chosen filter set facilitates accurate
photometric redshift estimation ($\sigma_z \approx 0.05$) reliable down to an $R$-band magnitude of 24.
During the observing runs, to facilitate accurate weak lensing studies, the best seeing conditions
were reserved for obtaining deep $R$-band images of the five fields. It is these $R$-band images,
along with the photometric redshift tables, that we make use of in this analysis. 


\subsection{Initial data processing}
The initial reduction of the data proceeded along the lines of that described in Gray et al. (2002, 
hereafter GTM+). The WFI instrument consists of a $4 \times 2$ array of $2048 \times 4096$ pixel CCDs. 
With a pixel scale of $0.238\arcsec$, the resulting total field of view (FOV) of the WFI is 
$0.56\degr \times 0.55\degr$. The standard COMBO-17 pre-processing pipeline (Wolf et al. in prep.)
produces mosaics of eight 2K $\times$ 4K chip images which have been debiased, corrected for non-linearity, 
normalized, flat-fielded and cleaned of cosmic rays. The resulting mosaics have been simply skewed 
as an approximate correction for rotational mis-alignments between the chips. We have removed this
skewing of the mosaiced images in order to restore the original chip configurations on the detector. 
The individual chip images were then extracted from the mosaics so that astrometric calibrations 
could be applied to each of the chip images individually.

\begin{figure*}
\vspace{3.0cm}
\centering
\begin{picture}(200,300)
\includegraphics{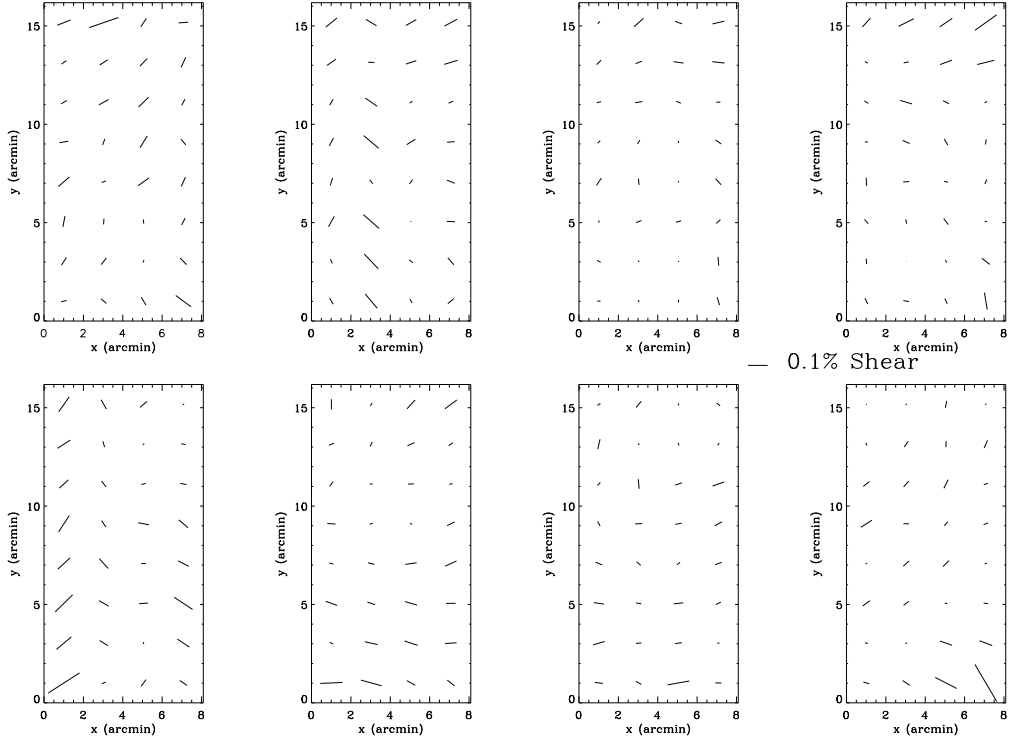}
\end{picture}
\vspace{-30mm}
\caption{Instrumental shear pattern for the 8 chips of the WFI instrument as measured from the observed
positions of the same objects on three dithered frames. The length of each vector represents the magnitude
of the instrumental shear at the indicated position on the chip and the direction indicates the
orientation of the distortion. For comparison, a $0.1\%$ shear is also shown on the same scale.}
\label{fig1}
\end{figure*}

\subsection{The precise astrometric solution}

After extraction from the mosaics, catalogues of objects were 
created for each chip exposure using SExtractor (Bertin \& Arnouts 1996). 
Using pointing information from the image headers, an approximate 
transformation was calculated to convert the SExtractor $(x,y)$ pixel 
coordinates of the objects to celestial coordinates $(\alpha,\delta)$ 
for each exposure. These objects were then matched with objects from 
the digitized SuperCOSMOS Sky Survey (SSS)\footnote{Full access to the 
SSS is available via the World Wide Web at URL {\tt http://www-wfau.roe.ac.uk}} 
(Hambly et al. 2001) to within a tolerance of 5\arcsec. The resulting coordinate 
pairs were then used to iteratively calculate a linear astrometric solution for 
each chip image with the optical axis being used as the tangent point for 
projection. Typically around 300 objects per chip image were used for the final 
astrometric fits which had rms residuals of $\sim 0.2 \arcsec$ (c.f. the pixel
scale of 0.238\arcsec/pixel). Including higher order terms and/or a radial 
distortion term in the astrometric solution has been investigated by GTM+. 
Although, in some cases, the fit was improved when a radial distortion term 
was included, they found no improvement for the instrument as a whole. 
Details concerning the artificial shear introduced by such a radial 
distortion are discussed in GTM+ who deduced a radial distortion of  
$\delta r/r \sim 0.025 \%$ and a resulting instrumental shear pattern 
with amplitude $\gamma < 0.0001$. This level of distortion is clearly 
negligible in comparison to a typical $\Lambda$CDM cosmic shear signal 
of $\gamma \sim 0.01$ and so we have used the simple linear solution to 
perform astrometry on all the chip images. Further tests of the artificial 
shear introduced by both the WFI instrument and the telescope optics are
described in the next section. Having produced linear astrometric fits for 
all the chip images in each of the five fields, these fits were then used 
to register the images to the same coordinate system. The images were thus 
aligned and combined, using 3$\sigma$ bad pixel rejection with weighting 
by exposure time and scaling by the median pixel value. All bad columns and 
pixels were removed during the co-addition procedure due to the large number 
of images combined. The five resulting 8K $\times$ 8K images were trimmed to 
remove the under-sampled edges of the fields resulting in five 8192 $\times$ 
8192 pixel (32.5\arcmin $\times$ 32.5\arcmin) images.

\subsection{Artificial shear introduced due to instrument and image co-addition}

We have investigated the level of distortions introduced by the 
telescope optics and the WFI instrument by comparing the location 
of the same objects on dithered chip exposures. As documented in 
Bacon et al. (2000, hereafter BRE), the respective positions of the 
same object on two dithered frames, $f$ and $f'$ can be expressed as
  \be
	\bmath{x}^{f'} - \bmath{x}^{f} \simeq {\bf \Psi}
	(\bar{\bmath{x}}^f-\bar{\bmath{x}}^{f'}),
  \label{dist}
  \ee
where $\bmath{x}^f$ ($\bmath{x}^{f'}$) is the position of the object 
on frame $f$ ($f'$), $\bar{\bmath{x}}^f$ ($\bar{\bmath{x}}^{f'}$) is 
the position of the centre of frame $f$ ($f'$) and $\bf \Psi$ is the 
distortion matrix, which may be expressed in terms of observables as
  \be
	{\bf \Psi} \equiv \left(
	\begin{array}{cc}
	\kappa+\gamma_1 & \gamma_2 \\
	\gamma_2        & \kappa-\gamma_1
	\end{array} \right).
  \ee
Here, $\kappa$ and $\gamma_i$ are the spurious convergence and shear 
introduced by the geometrical distortions. By measuring the positions 
of the same objects on three dithered frames, the resulting two equations 
of form (\ref{dist}) can be solved for $\kappa$ and $\gamma_i$. We have 
used this method to map the instrumental distortion across the 8 component 
chips of the WFI and the resulting shear pattern is shown in Fig. \ref{fig1} 
along with an indication of a $0.1\%$ shear signal. We find, in agreement with 
GTM+, that the induced shear due to telescope and instrument distortions is 
indeed negligible ($\gamma<0.2\%$) over the entire field compared to the 
typical weak lensing signal one would wish to measure.

\begin{figure}
\centering
\begin{picture}(200,220)
\includegraphics{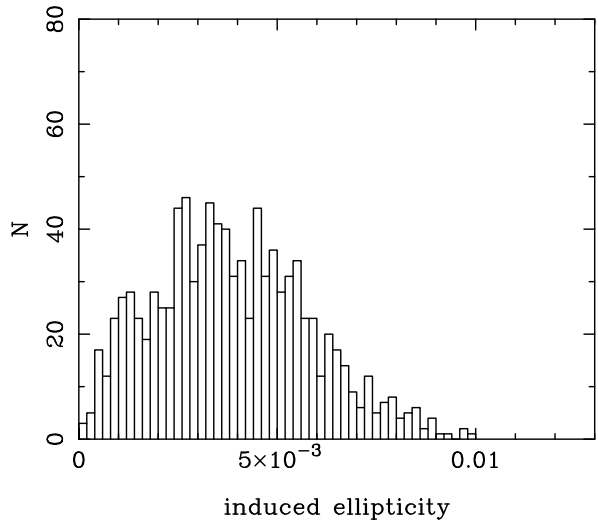}
\end{picture}
\vspace{-10mm}
\caption{Ellipticity introduced in a circular source due to co-addition of individual frames as
measured from 1000 simulations of the stacking of 78 frames.}
\label{fig2}
\end{figure}

We have also estimated the uncertainty introduced in the ellipticity 
measurements of objects on the combined images as a result of the co-adding 
procedure. Having registered the original exposures to the same coordinate 
frame, we then calculated the dispersion in the positions of many objects 
from the measured positions of those objects on all frames. We found a mean 
rms of $\approx0.5$ pixels in the positions of objects. Using this mean 
uncertainty in the position of objects, we can simulate the effect of 
image co-addition on a circular source of size $\sim0.7$ arcsec, similar to 
the mean seeing disc size on the individual frames. We simulate the stacking 
of 78 frames (c.f. no. of frames for CDFS field - see Table 1) where we apply 
to each image an offset, $\bmath{dx}=\bmath{dx}^{\rm true}+\delta\bmath{x}$ where
$\bmath{dx}^{\rm true}$ is the true offset position of the object and 
$\delta\bmath{x}$ is a vector whose magnitude is taken from a Gaussian 
distribution centred on $0.5$ pixels and whose orientation is taken from a 
uniform distribution between $0$ and $2\pi$. The size and ellipticity of 
the resulting stacked object was then computed using the formalism
described in section 5.3 of BRE. Fig. \ref{fig2} shows a histogram of 
the resulting ellipticity measurements for 1000 such simulations 
demonstrating that the shear arising from co-addition is again well 
below the sought-after cosmic shear signal. The median ellipticity 
induced in a circular source as measured from these simulations was 
$e_m=0.0037$ and the inter-quartile range was $\delta e_m=0.002$. These
numbers are well within our error budget for a precise measurement of 
cosmic shear. 

\subsection{Point spread function corrections and generation of
object catalogues}

We have used the {\tt imcat} software (Kaiser, Squires and Broadhurst, 
1995; hereafter KSB) to generate object catalogues for the five fields, measure shape 
parameters for all objects and to correct for such effects as the isotropic 
smearing of the point spread function (PSF) of objects by the atmosphere 
and telescope as well as any anisotropic smearing introduced by tracking 
errors, imperfect dither alignments and co-addition of the individual 
frames. We have used the {\tt hfindpeaks} routine to locate objects, 
followed by the {\tt getsky, apphot and getshapes} routines to estimate 
the local sky background, measure aperture magnitudes and half-light radii, 
$r_h$ and to calculate shape parameters for all detected objects 
(see e.g. KSB; BRE; GTM+, for details of how the KSB procedure and the
{\tt imcat} software works).

\begin{figure}
\vspace{-1mm}
\centering
\begin{picture}(200,150)
\includegraphics{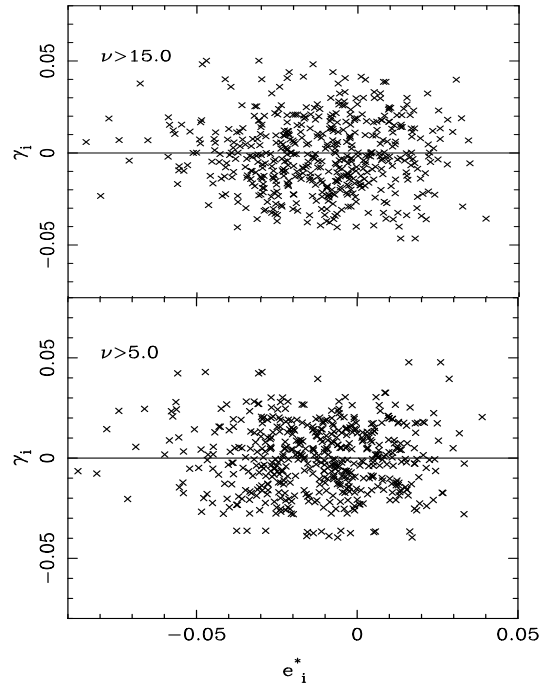}
\end{picture}
\vspace{40mm}
\caption{Mean shear components in cells plotted against the mean stellar 
ellipticity components for a cell size of $8\times8$ arcmin. The upper 
panel is for a S/N threshold of $\nu>15$ and the lower panel is for a 
threshold of $\nu>5$. The anti-correlation of $\gamma_i$ with $\overline{e_i^*}$ 
found in the analysis of BRE is not present in the COMBO-17 data 
for either of the S/N cuts.}
\label{fig3}
\end{figure}

This procedure produced an image catalogue for each field, containing positions, 
magnitudes, sizes and shapes for all the objects detected. We then applied 
reasonably conservative cuts on image size, signal-to-noise (S/N) and object 
ellipticity ($r_g>1.0, \nu>5, e<0.5$) to remove spurious and/or untrustworthy 
detections. Note that the S/N cut we have used ($\nu>5$) is much less 
stringent than that adopted in BRE who used a S/N cut of $\nu>15$. The 
motivation in BRE for using such a conservative cut was the existence in 
their data of a highly significant anti-correlation between their measured 
mean shear values and the corresponding mean stellar ellipticities for a S/N 
cut of $\nu>5$. We have searched for this effect in our data on several 
different smoothing scales and have found no evidence of correlations between 
our shear and stellar ellipticity values. For example, Fig. \ref{fig3} 
shows the mean shear components in cells plotted against the mean uncorrected 
stellar ellipticity components for a cell size of $8\times8$ arcmin. This is 
the same cell size as that shown in BRE's fig. 7 where a clear anti-correlation 
is present. It is clear from Fig. \ref{fig3} however, that this effect is 
absent from our data. We have looked for such an effect on scales ranging from 
$1$ to $30$ arcmin and have found none. We have therefore used a $\nu>5$ cut 
in our analysis. We also note that this is the same S/N cut as used in GTM+ for 
their supercluster analysis of the A901 field.

Having applied these cuts to the data, the {\tt imcat} software was then used 
to correct the measured shapes of the galaxies for the effects mentioned above. 
In particular, we correct the galaxy shapes for (a) anisotropy and (b) 
circularization. Details of the correction scheme used by {\tt imcat} have 
been described elsewhere (e.g. KSB; Luppino \& Kaiser 1997; 
Hoekstra et al. 1998; Kaiser et al. 1999) and will not be repeated here. 
Note that these corrections were done on sections (size $8' \times 16'$) of the 
fields rather than the full $32.5' \times 32.5'$ fields themselves. This was 
done to ensure that an accurate model of the PSF distortions was obtained. 
Performing the corrections on the full fields would possibly lead to 
residual PSF distortions in the dataset due to inaccurate modelling of the 
stellar ellipticities over such a large field.

\begin{figure}
\centering
\vspace{-5mm}
\begin{picture}(230,120)
\includegraphics{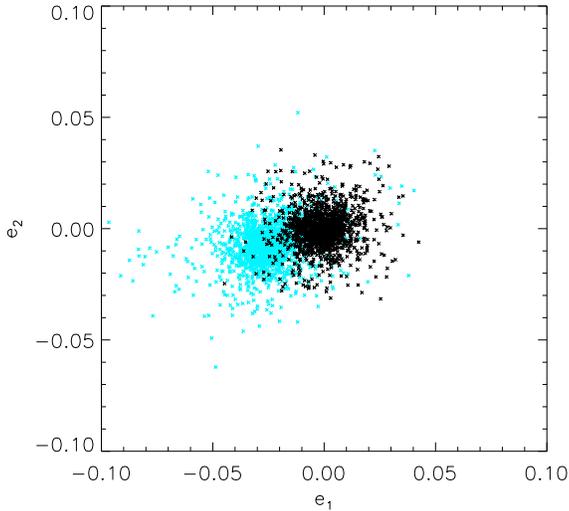}
\end{picture}
\vspace{30mm} \caption{Ellipticity distribution of stars in the
S11 field before (grey crosses) and after (black crosses) correction for PSF
anisotropy. Mean stellar ellipticity components of $\overline{e_1}=-0.025\pm0.013$ 
and $\overline{e_2}=-0.007\pm0.011$ are removed to leave an essentially randomly 
orientated stellar ellipticity distribution with mean residual 
components of $\overline{\delta e_1}=-0.0005\pm0.010$ and 
$\overline{\delta e_2}=-0.0002\pm0.009$.} \label{fig4}
\end{figure}

\begin{figure*}
\vspace{5mm}
\centering
\begin{picture}(200,300)
\includegraphics{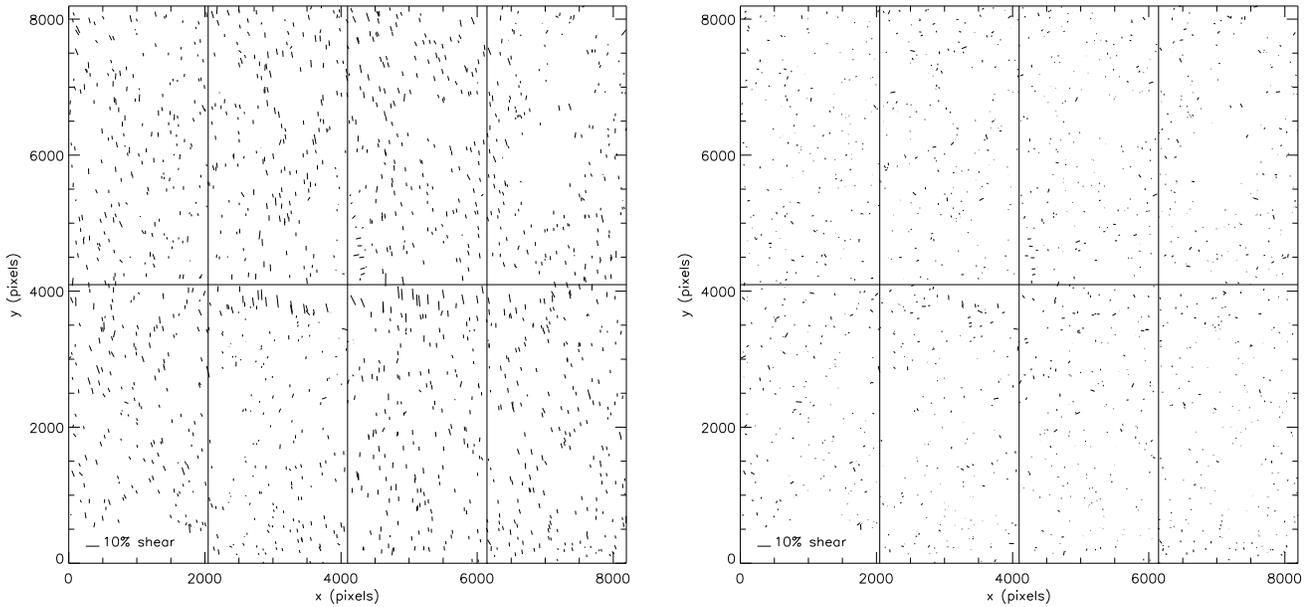}
\end{picture}
\vspace{-25mm} \caption{Stellar ellipticity pattern across the S11
field before ({\it left}) and after ({\it right}) correction for
PSF anisotropy. A $10\%$ shear signal is also shown for
comparison. The horizontal and vertical lines show the sections
used for the stellar ellipticity polynomial model fitting.}
\label{fig5}
\end{figure*}

\begin{table*}
\begin{minipage}{180mm}
\caption{Properties of the five COMBO-17 shear catalogues. In the sixth column we quote 
the median \rm{R}-band magnitude of each catalogue. In the case of the SGP and FDF fields, 
the median $m_{\rm R}$ listed is an approximate value as magnitudes for these fields have 
been roughly calibrated using the APM galaxy survey. The mean galaxy density values 
($n_{\rm gal}$) are calculated as $n_{\rm gal}=N_{\rm f}/A_{\rm f}$ where $N_{\rm f}$ and 
$A_{\rm f}$ are the total number of galaxies in the final shear catalogues and the total 
useable area of the fields respectively, after excluding regions contaminated by bright 
stars, diffraction spikes and ghosting (see Fig. \ref{shear_maps}). The FWHM values listed 
are the average FWHM of stars on each field as measured from the final co-added images. 
The final column indicates whether photometric redshift information is currently available 
for the field.}
\label{tab1}
\begin{tabular}{@{}lcccccccc}
\hline
Field & $\rm{RA}(2000)$ & $\rm{Dec.}(2000)$ & ${\rm R}$-band Exposures & Total Exp. Time & Median $m_{\rm R}$
      & $n_{\rm gal}$ (arcmin$^{-2}$) & FWHM & Redshift info.?\\
\hline
CDFS  & $03^{\rm{h}} \, 32^{\rm{m}} \, 25^{\rm{s}}$ & $-27\degr \, 48\arcmin \, 50\arcsec$
      & $42\times500 s+36\times420 s$ & $36120s$ & 24.4 & 37.5 & $0.81\arcsec$ & Yes\\
SGP   & $00^{\rm{h}} \, 45^{\rm{m}} \, 56^{\rm{s}}$ & $-29\degr \, 35\arcmin \, 15\arcsec$
      & $42\times500 s$ & $21000s$ & 23.6 & 36.2 & $0.81\arcsec$ & Pending\\
FDF   & $01^{\rm{h}} \, 05^{\rm{m}} \, 49^{\rm{s}}$ & $-25\degr \, 51\arcmin \, 42\arcsec$
      & $20\times500 s+5\times400 s$ & $12000s$ & 23.7 & 28.4 & $0.82\arcsec$ & No\\
S11   & $11^{\rm{h}} \, 42^{\rm{m}} \, 58^{\rm{s}}$ & $-01\degr \, 42\arcmin \, 50\arcsec$
      & $39\times400s +5\times500 s$ & $18100s$ & 23.7 & 27.7 & $0.78\arcsec$ & Yes\\
A901  & $09^{\rm{h}} \, 56^{\rm{m}} \, 17^{\rm{s}}$ & $-10\degr \, 01\arcmin \, 25\arcsec$
      & $36\times500 s+8\times600 s$ & $22800s$ & 24.0 & 36.1 & $0.76\arcsec$ & Yes\\
\hline
\end{tabular}
\end{minipage}
\end{table*}

A demonstration of the anisotropic PSF correction at work for the S11 field is given in
Figs. \ref{fig4} and \ref{fig5}. The smoothly varying stellar ellipticity pattern
apparent in the stars before correction (Fig. \ref{fig5}, left-hand side) is successfully 
removed to produce residual stellar ellipticities with essentially random orientations 
(Fig. \ref{fig5}, right-hand side) and mean residuals of $|\delta e_1|\approx5\times10^{-4}$ 
and $|\delta e_2|\approx2\times10^{-4}$. Fig. \ref{fig4} shows the stellar ellipticities 
after correction randomly distributed about $e_1=e_2=0$. Note however that these plots are 
for illustrative purposes only. Rigorous tests of the systematics introduced by residual 
spurious ellipticities are performed later in Section 5, where we measure the star-galaxy 
cross-correlation functions for the dataset. 

\begin{figure*}
\vspace{3.0mm}
\centering
\begin{picture}(200,200)
\includegraphics{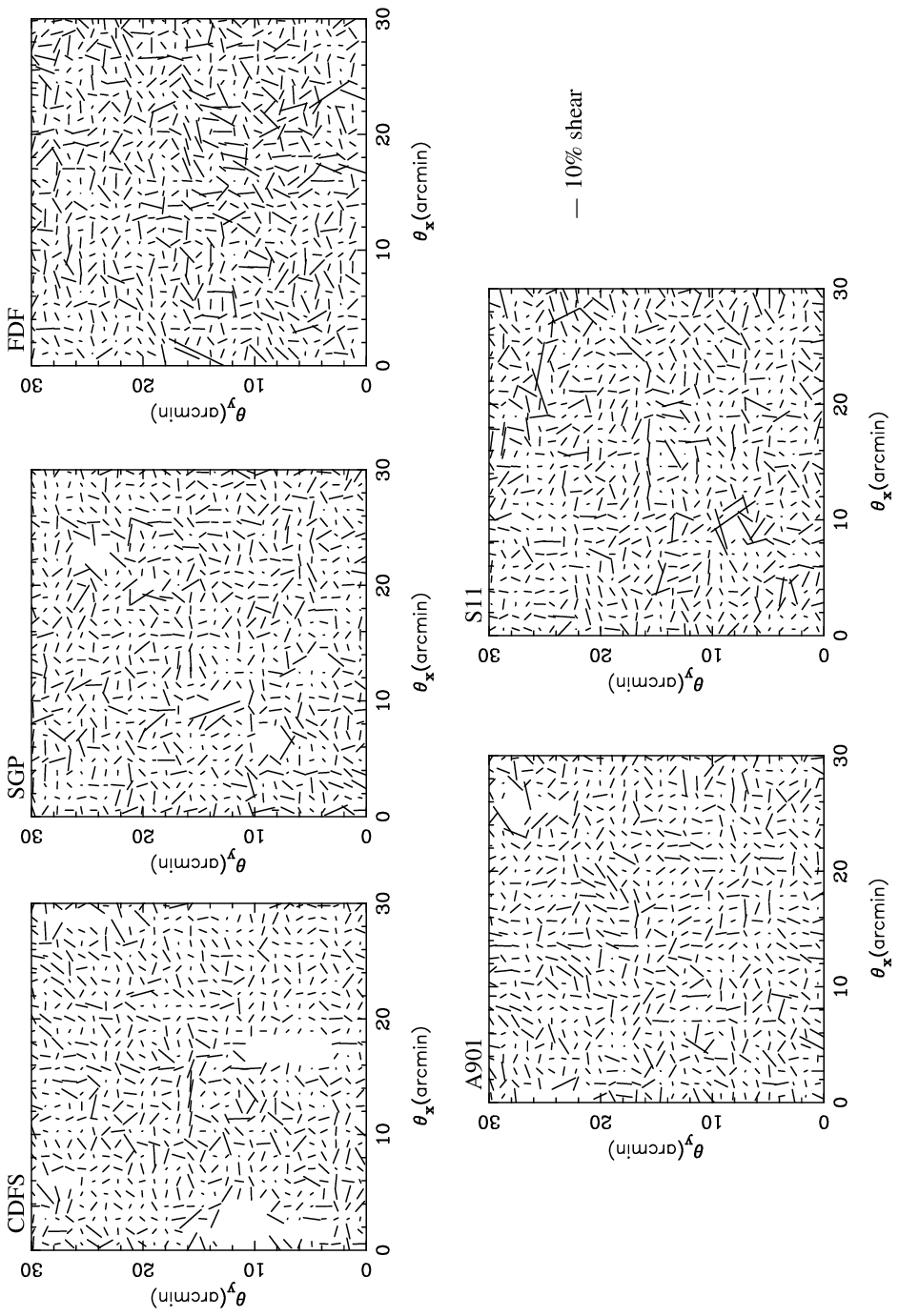}
\end{picture}
\vspace{25mm}
\caption{The shear distribution on the five COMBO fields. The fields are 
(clockwise from top left) CDFS, SGP, FDF, S11 and A901. The shear measurements 
for each field have been binned into $30 \times 30$ pixels, giving a pixel scale 
of $\sim 1$ arcmin. For each pixel, the length of the vector drawn is proportional 
to the magnitude of the mean shear in that pixel and its direction indicates the 
orientation of the shear. A $10\%$ shear signal is also shown for comparison. The 
apparent holes in the distributions are regions which have been masked out due to 
bright stars, diffraction spikes etc.} 
\label{shear_maps}
\end{figure*}

\section{The COMBO-17 fields}
\subsection{Content of the fields}
The five fields observed are quite different from one another in terms of their content. 
Three are blank fields; one (CDFS) was chosen so it would overlap with the Chandra Deep 
Field South; another (SGP) is centred on the South Galactic Pole, and the third blank 
field is the shallower FORS deep field (FDF). A fourth, randomly selected field (S11) contains a 
fairly large cluster (Abell 1364) at a redshift of $z=0.11$. The fifth field (A901) was 
chosen so as to include a supercluster system (Abell 901/902) at redshift $z=0.16$ and 
the COMBO-17 observations of this field have already been the subject of an extensive weak 
lensing analysis, details of which are given in GTM+. Details of the $R$-band 
observations for the five fields are given in Table \ref{tab1}. Note that we do not currently
have the full 17-band photometric information for the SGP or FDF fields. 

The significant mass concentrations present in the two fields containing clusters 
(S11 \& A901) raise the question as to whether these fields should be included in a cosmic 
shear analysis at all. The significant difference between these fields is the pre-selection 
of A901 to include a supercluster system. Strictly speaking, therefore, the A901 field is not 
a randomly selected piece of sky and is unlikely to be a fair representation of the Universe. 
We have therefore chosen not to include this field in our final parameter estimation and shear 
power spectrum calculations. We have also experimented with including/excluding both the S11 and  
A901 fields from the analysis. For both the shear power spectrum reconstruction and the parameter 
estimation, we find that the A901 field biases our results significantly, whereas the S11 field 
is consistent with the dataset as a whole (see Section 6.3). The cell-averaged shear distributions 
for the five fields are shown in Fig. \ref{shear_maps}, showing regions in the fields which have 
been masked out due to contamination by bright stars, diffractions spikes and ghosting.  

\subsection{Including redshift information}

As mentioned earlier, the COMBO-17 survey has the unique
advantage, in terms of weak lensing surveys, of having accurate
photometric redshift measurements for $\sim90\%$ of objects detected 
down to a magnitude of $m_R \leq 24$ (for a detailed description 
of the photometric methods used to assign redshifts to the galaxies, 
see Wolf et al. 2003). For these galaxies, the estimated typical 
uncertainty in redshift is $\sigma_z \sim0.05$. At the time of writing, 
photometric data is available for three of our five fields (CDFS, A901 \& S11). 
We have used this information to estimate, directly from the data, the redshift 
distribution and in particular, the median redshift, $z_m$ of the lensed source 
galaxies in the survey. For the $z_m$ calculation, we have only included those 
galaxies used in the final weak lensing analysis. This subset of galaxies does, 
in fact, include galaxies with magnitudes $m_R >24$ for which the redshift 
estimations are unreliable. To account for this, we must extrapolate 
the redshift distribution beyond $m_R=24$. For the three fields for which 
photometric data is available, accurate redshifts have been found for $31649$ out 
of the $83514$ galaxies used in the final shear analysis ($\sim38\%$).  

\begin{table}
\center
\caption{Measured median magnitudes and redshifts ($z_m$) for the three fields, 
CDFS, A901 \& S11 for different limiting magnitude cut-offs. Also listed are the
number of galaxies ($N_{\rm gal}$) used in the median calculations and the 
completeness of each magnitude-limited sample (i.e. the fraction of galaxies in the
sample for which accurate redshifts have been obtained). Extrapolating the median 
$m_{\rm R}$ -- $z_m$ relation (see Fig. \ref{mag-z}) to the median magnitude of 
the combined shear catalogues for all fields ($=24.0$), we infer a median 
redshift for the COMBO-17 weak lensing survey of $z_m=0.85\pm0.05$.}
\label{tab2}
\begin{tabular}{@{}lccccc}
\hline
Limiting $m_{\rm R}$ & $N_{\rm gal}$ & Median $m_{\rm R}$ & $z_m$ & 
Completeness \\
\hline
$<20.0$ & $1485$  & $19.15$ & $0.184\pm0.004$ & $87\%$ \\
\\
$<21.0$ & $3267$  & $20.13$ & $0.285\pm0.030$ & $87\%$ \\
\\
$<22.0$ & $7316$  & $21.14$ & $0.404\pm0.026$ & $88\%$ \\
\\
$<23.0$ & $16387$ & $22.14$ & $0.520\pm0.023$ & $87\%$ \\
\\
$<24.0$ & $31649$ & $22.96$ & $0.673\pm0.092$ & $76\%$ \\

\hline
\end{tabular}
\end{table}

For a number of different limiting magnitudes, we have measured the median 
magnitude of the galaxies and the corresponding median redshift ($z_m$) of 
those same galaxies. Details of these measurements are given in Table 
\ref{tab2}. We take our median redshift measurements as lower limits for the 
true median redshift as the magnitude-limited samples are not complete in redshift 
(see Table 2) and those galaxies without an assigned redshift are most likely to be
at a higher redshift than our measured median value. To calculate an upper limit for 
our $z_m$ estimates, for each magnitude limited sample, we have placed those galaxies 
without redshift measurements at $z=\infty$ and have re-calculated $z_m$. We take our 
final median redshift estimate to be simply the midpoint between our upper and lower limits. 

\begin{figure}
\centering
\begin{picture}(200,250)
\includegraphics{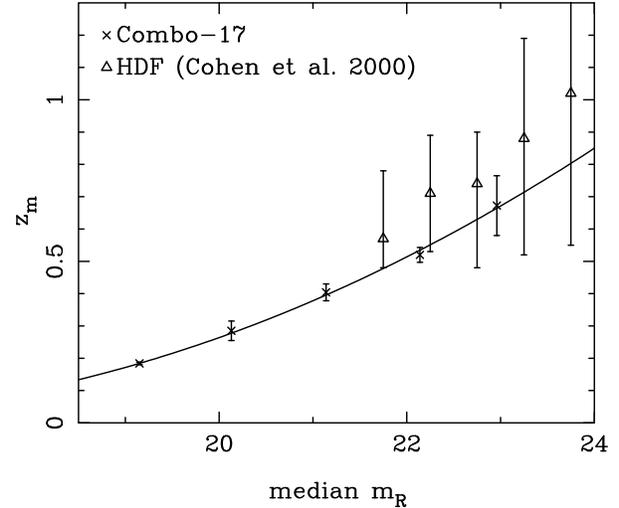}
\end{picture}
\vspace{-12mm}
\caption{The dependence of the median redshift ($z_m$) of the COMBO-17 galaxies on the 
median magnitude of the galaxies as measured from the 3 fields for which we have 
photometric data. A simple quadratic model (solid curve) for the $m_{\rm R}-z_m$ relation is also 
plotted for comparison. For our final weak lensing dataset, which has a median 
{\rm R}-band magnitude of $m_{\rm R}=24.0$, we infer from this trend a median 
redshift of $z_m=0.85 \pm 0.05$.} 
\label{mag-z}
\end{figure}

We have then extrapolated the median $m_{\rm R}$ -- $z_m$ relation to find an 
estimate of the median redshift of our galaxy sample as a whole, which has a 
measured median magnitude of 24.0. The median $m_{\rm R} - z_m$ relation is 
shown in Fig. \ref{mag-z}. Here we plot both the COMBO-17 data points and data
from the HDF redshift survey of Cohen et al. 2000, with which we are fully 
compatible. The curve plotted is the best-fit quadratic model to the COMBO-17 data
and is given by
\be
z_m=2.53-0.33 \,m_{\rm R,m}+0.01 \,m_{\rm R,m}^2.
\ee
where $m_{\rm R,m}$ is the median $R~$band magnitude of the galaxy sample. From this, we 
estimate a median redshift of $z_m=0.85 \pm 0.05$ for the COMBO-17 weak lensing survey. 

There is, of course, more information in the measured redshift 
distribution than just the median value. The advantage of including
photometric information in a weak lensing analysis lies in reducing or 
eliminating uncertainties in the redshift distribution of the source 
galaxies when it comes to comparing weak lensing measurements with those 
predicted from theory (see Sections 6 \& 7). To this end, we have included 
the measured redshift distribution for our comparison with theoretical models. 
There are, of course, many more galaxies for which we have no redshift 
estimate than there are galaxies with reliable measurements. 
This situation will improve as the COMBO-17 survey nears completion, 
but for the purposes of this analysis and predicting the expected weak 
lensing signal for different cosmological scenarios, we need to assign 
redshifts to the galaxies with unknown redshifts. We do this by distributing 
these galaxies in $z$ according to (Baugh \& Efstathiou 1991) 
\be
\frac{dN}{dz}=\frac{\beta z^2}{z^3_{\star}} \exp \left[ 
-\left( \frac{z}{z_{\star}} \right)^{\beta} \right],
\label{nofzeqn}
\ee
with $\beta=1.5$ and $z_{\star}=z_m/1.412$. A distribution of the form, equation 
(\ref{nofzeqn}) has a median redshift of $z_m$. We have tuned the value of $z_m$
in equation (\ref{nofzeqn}) to ensure that the final \emph{total} redshift distribution
has a measured median value of $z_m=0.85$.   

\begin{figure}
\centering
\begin{picture}(200,250)
\includegraphics{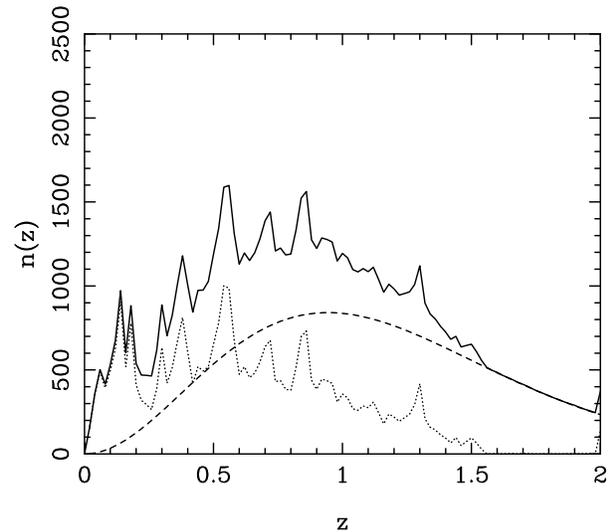}
\end{picture}
\vspace{-12mm}
\caption{The final redshift distribution (solid line) used for making model predictions 
to compare with the observed weak lensing signal from the COMBO-17 dataset. This 
composite distribution has a median redshift of $z_m=0.85$ and is composed of a measured 
$n(z)$ distribution (from the three fields, CDFS, S11 and A901; shown here as the dotted 
line) and a parameterised model (equation (\ref{nofzeqn}); dashed line) for the galaxies 
with unknown redshifts.} 
\label{nofz_total}
\end{figure}

After assigning values to the galaxies with 
unknown redshifts according to equation (\ref{nofzeqn}), we now have the final 
$n(z)$ distribution which we can use for making predictions for the weak lensing 
signal expected in the COMBO-17 survey. This final $n(z)$ distribution is shown 
in Fig. \ref{nofz_total}.  
	
\section{Correlation functions and shear variance estimators}

It is useful to estimate the correlation functions of the shear field both as a
first approach at measuring the strength of the cosmic shear signal,
but also as a test for unwanted systematics in the data. These
tests have now become a standard part of shear analysis (e.g. Van Waerbeke et al. 2001; 
Pen, Van Waerbeke \& Mellier 2002; Bacon et al. 2002). Measuring 
correlation functions also allows a direct comparison of our measurements with
previous cosmic shear studies. Here we present both an unweighted correlation function 
analysis and a minimum variance weighted shear variance analysis. Note that, for the 
remainder of the paper, we have excluded the A901/2 supercluster field from our 
measurements except where specifically indicated.  

\subsection{Correlation analysis}

To estimate the unweighted correlation functions, we follow Bacon et al. (2002).
For the purposes of the correlation analysis, we have divided each of the COMBO-17 
fields into eight, chip-sized sections. This allows for a better estimate of the 
field-to-field covariance between our measurements than would be possible using the 
original $0.5\degr \times 0.5\degr$ fields. The correlation functions for the 
individual sections can be measured by averaging over galaxy pairs:
 \ba
        C_{1,s}(\theta) & = & \frac{1}{n_s}\sum_{i=1}^{n_s}
        \gamma_1^r(\rhat_i)\gamma_1^r(\rhat_i+\thetab), \nn
        C_{2,s}(\theta) & = & \frac{1}{n_s}\sum_{i=1}^{n_s}
        \gamma_2^r(\rhat_i)\gamma_2^r(\rhat_i+\thetab), \nn
        C_{3,s}(\theta) & = & \frac{1}{n_s}\sum_{i=1}^{n_s}
        \gamma_1^r(\rhat_i)\gamma_2^r(\rhat_i+\thetab), 
 \ea
where $n_s$ is the number of pairs of galaxies in each section 
and where we average over all galaxy pairs separated by an 
angle, $\theta$ (c.f.  equation (\ref{corrfns})). The signal 
averaged over all sections is simply
 \be
        \overline{C}_i(\theta) = \frac{1}{N_s}
        \sum_{s=1}^{N_s} C_{i,s}(\theta) , 
        \label{corr}
 \ee
where $N_s$ is the total number of sections. The covariance of the 
correlation function measurements is given by  
 \ba
        && {\rm cov}[\overline{C}_i(\theta) \overline{C}_j(\theta')] \simeq \nn
        && \,\,\,\,\,\,\,\,  \frac{1}{N^2_s}\sum_{s=1}^{N_s}
          [C_{i,s}(\theta)-\overline{C}_i(\theta)]
          [C_{j,s}(\theta')-\overline{C}_j(\theta')].
        \label{covcorr}
 \ea 

\begin{figure}
\centering
\begin{picture}(200,230)
\includegraphics{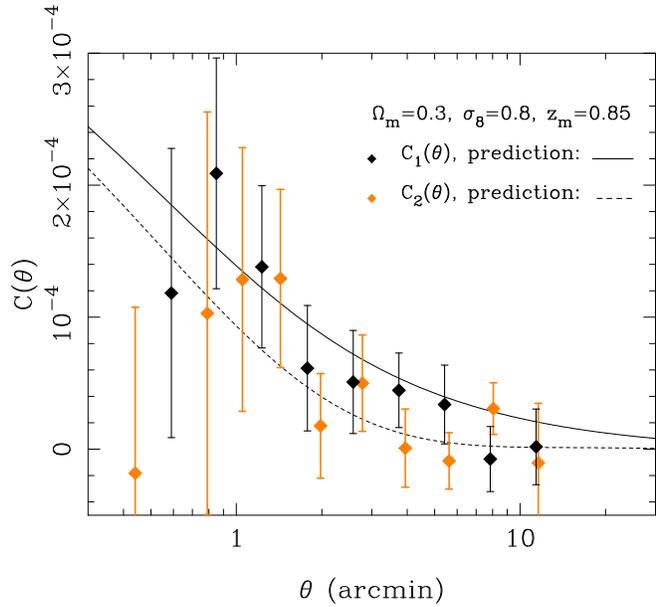}
\end{picture}
\caption{The unweighted shear correlation functions $C_1(\theta)$
and $C_2(\theta)$ from the COMBO-17 data. The solid dark line is
the expected $C_1$ correlation for a $\Omega_m=0.3 \,\, \Lambda$CDM model, 
normalised to $\sigma_8=0.8$, while the lighter dashed line is the 
expected $C_2$ signal. The $C_2$ points have been slightly displaced 
horizontally for clarity. Note that we measure an inconsistently large 
$C_1$ signal ($>5 \times 10^{-4}$) on the very smallest 
scales ($\theta<0.3$ arcmin, not shown) which, we suspect is due to 
residual systematic effects and/or intrinsic alignments.} 
\label{c1c2}
\end{figure}

\begin{figure}
\centering
\begin{picture}(200,230)
\includegraphics{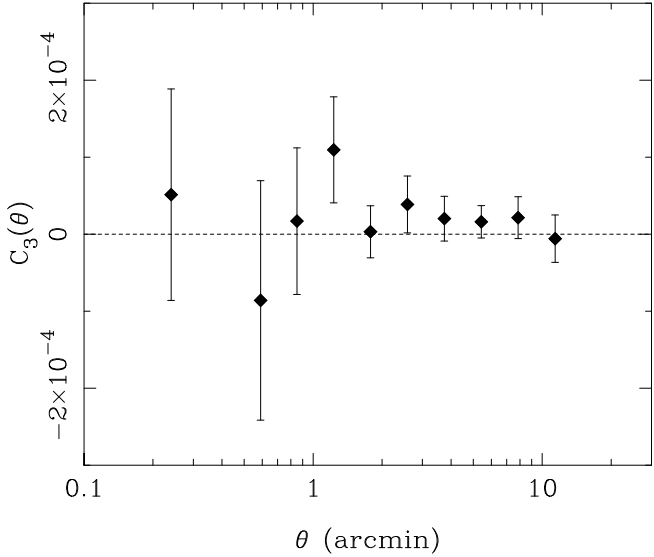}
\end{picture}
\caption{The cross-correlation of the shear components $C_3(\theta)$, as measured 
from the COMBO-17 dataset. The signal is consistent with zero on all scales in 
agreement with theoretical predictions.} \label{c3}
\end{figure}

\begin{figure}
\centering
\begin{picture}(200,230)
\includegraphics{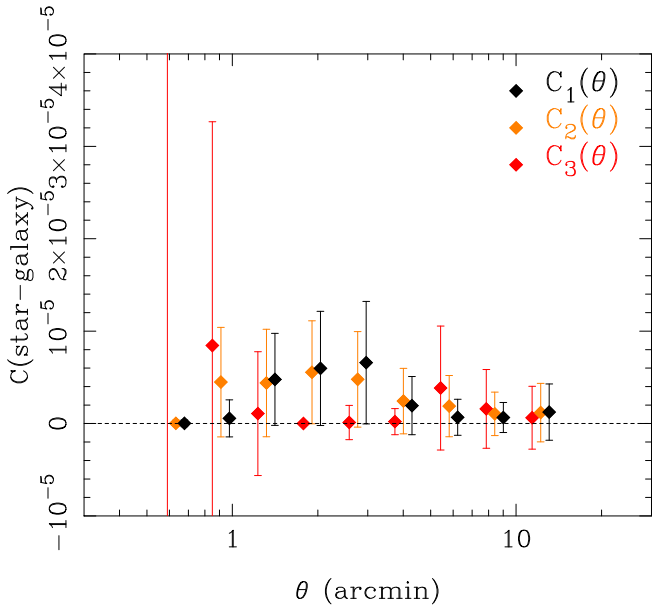}
\end{picture}
\caption{The star-galaxy cross-correlation functions (equation (\ref{star_cont})) 
for the COMBO-17 data. The $C_1$ and $C_2$ points have been slightly offset 
horizontally for clarity. These measurements are a strong test of how successfully the 
galaxies have been corrected for distortions introduced by PSF anisotropy. 
The residual correlations between galaxies and stars are consistent with zero on all 
scales. In the worst-case scenario, represented by the upper end of the error bars shown, 
our correlation function measurements would include $<\!10\%$ contamination from residual 
distortions left over from the PSF corrections applied.}
\label{star-gal}
\end{figure}

Fig. \ref{c1c2} shows the correlation functions, $C_1(\theta)$ and $C_2(\theta)$ 
after averaging over all sections of the four fields, CDFS, S11, SGP and FDF. 
As can be seen, our data agrees well with the expected correlations, 
for both $C_1$ and $C_2$, for a $\Omega_m=0.3 \,\,\, \Lambda$CDM model 
with a power spectrum normalisation of $\sigma_8=0.8$ (see Section 7 
for our measurements of $\Omega_m \,\, \& \,\,\sigma_8$ from COMBO-17). 
These theoretical curves have been calculated using equation (\ref{corr2cl}) 
where we have used the halofit model of Smith et al. 
(2002)\footnote{the halofit code is publicly available at URL 
\tt{http://www.as1.chem.nottingham.ac.uk/$\sim$res/software.html}}  
to calculate the non-linear matter power spectrum. We have also input our 
combined redshift distribution (Fig. \ref{nofz_total}) into the calculation. 
We should point out that on smaller scales than those shown on Fig. \ref{c1c2}, 
i.e. $\theta \la 0.3$ arcmin, we have measured an inconsistently large signal 
($\ga 5\times10^{-4}$) for the $C_1$ correlation function. We suspect that this 
greatly enhanced signal is probably due to either systematic effects on very 
small scales due to imperfections in the correction for PSF anisotropy, or 
alternatively, the increased signal may be caused by intrinsic alignments dominating 
at such small angular separations. Either way, we exclude this single data point 
for our parameter estimations in Section 7. We also note that the errorbars on 
$C_1(\theta)$ and $C_2(\theta)$ may be slightly underestimated on the very largest 
scales due to correlations between neighbouring sections of the same field. 
Fig. \ref{c3} shows the cross-correlation between shear components, $C_3(\theta)$ 
which is consistent with zero, as expected.

In Fig. \ref{star-gal}, we show the star-galaxy cross-correlations 
for our data. The systematically-induced cross-correlation between 
the galaxies (corrected for PSF anisotropy) and the uncorrected stars 
is defined by (Bacon et al. 2002):
 \be
	C_i^{\rm sys}=\frac{\lgl \gamma_i e^*_i \rgl^2} 
	{\lgl e^*_i e^*_i \rgl},
 \label{star_cont}
 \ee
where $i=1,2$. A correlation here would indicate a 
problem in the correction of systematic image distortions. Again we 
find that no correlations are detected at scales $\theta\ga0.3$ arcmin, 
indicating that our PSF correction procedure has worked well. Note, however 
that the star-galaxy cross-correlations are significant at scales $\theta<0.3$ 
arcmin, indicating a possible systematic problem at these very small scales -- 
this motivates our exclusion of the smallest-scale $C_1(\theta)$ data point for
our parameter estimations in Section 7. 

\begin{figure}
\centering
\begin{picture}(200,230)
\includegraphics{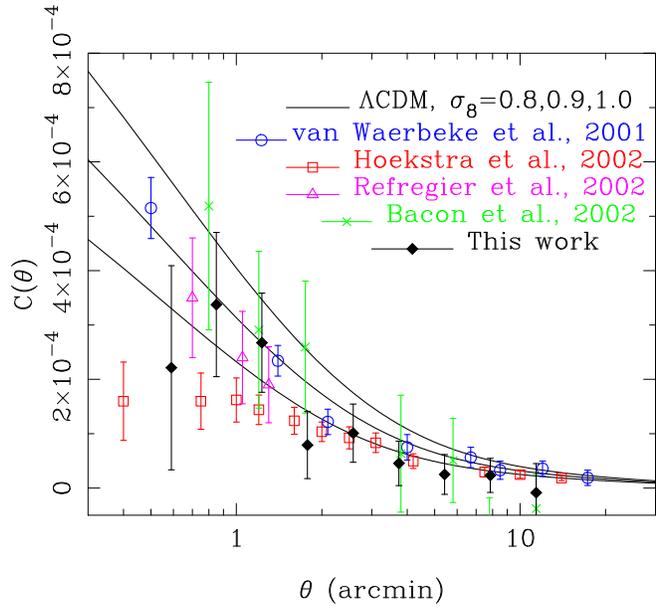}
\end{picture}
\caption{The total correlation function, $C(\theta)$, as measured from COMBO-17, along with 
the most recent cosmic shear measurements from the four other groups indicated. Beyond a scale 
of 1 arcmin, the measurements are in broad agreement. The correlation function predictions for
a flat $\Lambda$CDM cosmology, for three values of the power spectrum normalisation 
(from top to bottom: $\sigma_8=1.0,0.9,0.8$) are also plotted.} 
\label{comparison}
\end{figure}

\begin{figure}
\centering
\begin{picture}(200,230)
\includegraphics{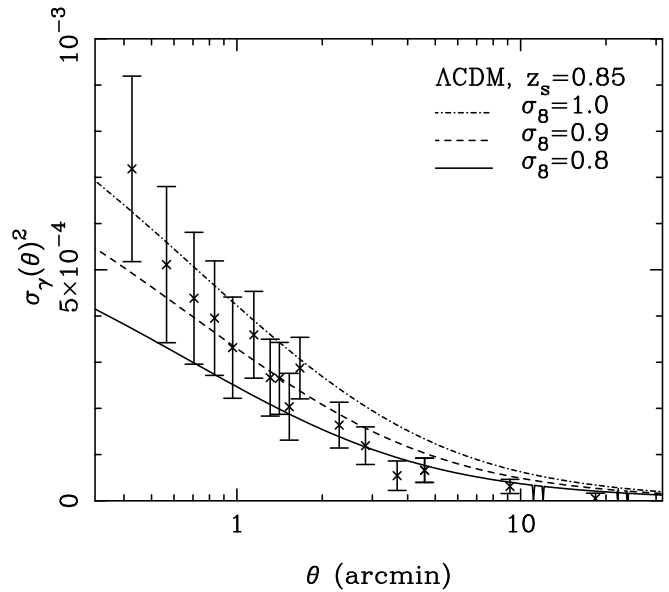}
\end{picture}
\caption{The minimum variance shear variance as measured from the COMBO-17 
dataset. Also plotted are predictions for the shear variance for a $\Omega_m=0.3 
\,\, \Lambda$CDM cosmology for three different normalisations of the matter power 
spectrum.} \label{shearvar}
\end{figure}

We can use our correlation function measurements to compare our results with those of 
previous cosmic shear studies. We perform such a comparison by looking at the 
total correlation function, $C(\theta)=C_1(\theta)+C_2(\theta)$ for each data set. In
Fig. \ref{comparison}, we plot $C(\theta)$ as measured from COMBO-17, along with the 
most recent results from four other groups (Van Waerbeke et al. 2001; Bacon et al. 2002; 
Refregier et al. 2002; Hoekstra et al. 2002). Note that we have scaled each groups' 
results to a median redshift of $z_s=0.85$ using the scaling suggested by the numerical 
simulations of Barber (2002), i.e. $C(\theta) \propto z_m^2$. We have also, in the case of 
Refregier et al., and  Hoekstra et al., assumed that $\sigma_{\gamma}^2(\theta) \approx C(\theta)$, 
where $\sigma_{\gamma}^2(\theta)$ is the shear variance statistic measured by these two groups.
That this is a reasonable assumption can be seen by comparing the $\sigma_8=0.9$ models in Figs. 
\ref{comparison} and \ref{shearvar}. Fig. \ref{comparison} shows that, beyond a scale of $\theta\sim1$ 
arcmin, all groups are in broad agreement with each other, although the COMBO-17 signal does 
seem to be slightly lower in amplitude. At scales smaller than 1 arcmin, comparisons are more 
difficult to make because of the larger error bars involved. Indeed, it is the measurements 
at the larger scales ($\theta \ga 1$ arcmin) that provide our best constraints on 
cosmological models.   

\subsection{Shear variance}

We have also applied the minimum variance estimator of Brown et al.
(2002) to measure the shear variance in square cells in excess of
the noise:
 \be
    \sigma^2_{\rm lens}(\theta)=\frac{ \sum_{\rm cell}  w_{\rm cell}
                (e^2_{\rm cell}-N_{\rm cell})}
               {\sum_{\rm cell} w_{\rm cell}},
    \label{mv}
 \ee
where $e_{\rm cell}$ is the cell averaged ellipticity and $w_{\rm cell}$
is a minimum variance weight;
 \be
    w_{\rm cell} = \frac{1}{2[\sigma^4_{\gamma}(\theta) + N^2_{\rm
    cell}(\theta)]}.
 \ee
Here $N_{\rm cell}$  is the noise in the cell measured from the data
due to the intrinsic dispersion in galaxy ellipticities,
 \be
        N_{\rm cell} (\theta) = \frac{1}{n^2} \sum_{i=1}^n
        \left(e^2_1(\rhat_i) + e^2_2(\rhat_i)\right),
 \ee
where $n$ is the number of galaxies in a cell
and $\sigma^2_{\gamma}$ is the predicted shear variance
for square cells which we calculate for a $\Lambda$CDM model using 
$\sqrt{\pi} \, \times$ equation (\ref{var2cl}). 
Varying the cell side length yields the cell variance as a function of scale, 
$\sigma^2_{\rm lens}(\theta)$. The error in equation (\ref{mv}) is given by
 \be
      {\rm Var}[\sigma^2_{\rm lens}]= \frac{1}{ \sum_{\rm cell} w_{\rm cell}} .
       \label{errmv}
 \ee

In Fig. \ref{shearvar}, we plot our shear variance measurements along with the 
predicted shear variance signal for a $\Lambda$CDM model, and for three values of the
power spectrum normalisation, $\sigma_8=0.8,0.9 \, \, \& \, 1.0$. Immediately, one 
sees that the shear variance statistic is measuring a somewhat higher signal 
than the correlation function on scales $\la 2$ arcmin. We suspect that this is the
same effect as that seen in the correlation functions at very small scales, i.e.
that at these scales residual systematics and/or intrinsic galaxy alignments 
become important. Very recently, methods have been developed for separating the 
intrinsic and lensing signals (Heymans \& Heavens 2003; King \& Schneider 2002) 
by effectively down-weighting physically close pairs of galaxies when calculating the 
correlation functions. Such methods are dependent on accurate photometric information 
being available for the individual galaxies. In this respect, the COMBO-17 survey is 
ideal for the separation of the intrinsic and lensing signals, and we are currently
applying such an analysis to the survey. 
We note, for the purposes of the current lensing analysis, however, that intrinsic 
alignments are unlikely to contribute significantly to our signal beyond scales of a 
few arcmin in the shear variance or beyond $\sim$ 1 arcmin in the correlation functions.
Indeed, the few observational studies of intrinsic alignments carried out to date  
(Pen, Lee \& Seljak 2000; Brown et al. 2002; Lee \& Pen 2002) support arguments from 
both theoretical considerations (Crittenden et al. 2001; Catelan et al. 2001; 
Mackey et al. 2002) and numerical simulations (Heavens, Refregier \& Heymans 2000; 
Croft \& Metzler 2001) that predict the intrinsic contribution to be no more than 
$\sim10\%$ for deep surveys such as COMBO-17. We do, however, urge caution when 
interpreting these statistics on scales $\la 1$ arcmin.

\section{Cosmic Shear Likelihood Analysis}
Having tested our data for sources of systematic errors and measuring 
the cosmic shear signal by standard methods, we now apply a new maximum 
likelihood analysis to estimate the cosmic shear power spectrum from 
the COMBO-17 dataset. This section contains the key results of our
paper.

\subsection{Likelihood procedure}
Our approach is based on the prescription of Hu \& White
(2001, hereafter HW), who use a maximum likelihood method to
reconstruct the three power spectra, $C^{\kappa\kappa}_\ell$,
$C^{\kappa \beta }_\ell$ and $C^{\beta\beta}_\ell$, directly from
pixelised shear data. More precisely HW proposed reconstructing
$C^{\kappa\kappa}_{\ell}$ as a series of step-wise ``band powers'',
extracted from the data via an iterated quadratic estimator of 
the maximum likelihood solution. This approach, which is similar 
to methods used in analyses of CMB polarisation fields 
(e.g. Tegmark \& de-Oliviera Costa 2002), has the advantage that it
automatically accounts for irregular survey geometries and
produces error estimates which include sampling variance and shot
noise. In addition this approach can account for the effects of
pixelisation. 

HW have tested the maximum likelihood estimator on both 
Gaussian realisations of a $\Lambda$CDM power spectrum and on N-body
simulations. In both cases, the estimator performs well,
recovering the input power accurately with error estimates from
the Fisher matrix showing excellent agreement with run-to-run
errors. We also investigate the method with our own simulations in 
the next section. 

Another advantage of this approach for reconstructing the shear 
power spectrum is that it provides a simple method for performing a
decomposition of the signal into curl ($\beta$) and curl-free 
($\kappa$) modes. The weak lensing shear power spectrum is 
predicted to be completely curl-free in the absence of significant 
lensing from gravitational waves (Stebbins 1997; Kamionkowski et al. 1998)
although the predictions for intrinsic galaxy alignments are less
certain (e.g. Crittenden et al. 2001, Mackey et al. 2002).
The $\kappa$/$\beta$ decomposition, therefore, represents a useful 
method for the detection of non-lensing artefacts (e.g. intrinsic 
alignments, systematic effects) in the data.

If we write our data as a vector, 
 \[
	\db = (\gamma_1(\rhat_1),\gamma_2(\rhat_1),\cdots,
	\gamma_1(\rhat_n),\gamma_2(\rhat_n)), 
  \]
then the likelihood function is
 \be
   -2 \ln L(\C|\db) = \db^t \C^{-1} \db + {\rm Tr} \, [{\rm ln} \C],
 \ee
where
 \be
        \C = \lgl \db \db^t \rgl
 \ee
is the data covariance matrix, and we assume uniform priors. 
We can interpret the data covariance matrix as the sum of the 
shear covariance matrix (equation (\ref{shearcovar})) and a noise term
 \be
	\N=\frac{\gamma^2_{\rm rms}}{N_{\rm pix}}\I,
 \label{noisematrix}
 \ee	
which we measure directly from the data. Here, $\I$ is the identity 
matrix, $\gamma_{\rm rms}$ is the intrinsic dispersion of galaxy 
ellipticities within a pixel and $N_{\rm pix}$ is the pixel occupation 
number.   

Following HW, we maximise the likelihood as a function of 
the model parameters. Here, our model parameters are just the 
band powers of the three power spectra, $C^{\kappa\kappa}_{\ell},
C^{\beta\beta}_{\ell}$ and $C^{\kappa\beta}_{\ell}$ and we perform
the maximization iteratively with a Newton-Raphson scheme.
That is, from an initial guess of the band powers, 
$\theta_i$, a new estimate,
 \be 
	{\theta_i}'=\theta_i+\Delta\theta_i
 \ee
is made for the band powers where we have adjusted our previous 
estimate by
 \be 
	\Delta\theta_j=2 \left(\frac{\partial\,{\rm ln}L}
         {\partial \theta_i}\right) \left( \frac{\partial^2\,{\rm ln} L}
         {\partial \theta_i \partial \theta_j} \right)^{-1}.
 \ee  
Here, we can replace the second derivative of the likelihood by its 
expectation value,
 \be
	\left< \left(\frac{\partial^2\,{\rm ln} L} {\partial 
        \theta_i \partial \theta_j} \right)^{-1} \right> \approx 
        F^{-1}_{ij},
 \ee 
and at the same time, in the limit where the likelihood is sufficiently 
Gaussian in the parameters, we can use the Fisher Information matrix 
(e.g. Tegmark, Taylor \& Heavens 1997),
  \be
    F_{ij} = \frac{1}{2} {\rm Tr}\, [\C^{-1} \de_{\theta_i} \!\C]
    \,.
    [\C^{-1} \de_{\theta_j}\! \C]
 \ee
to estimate the uncertainties on the band power measurements. That is, we 
approximate the covariance matrix of our band power estimates as 
 \be
   {\rm cov}[\theta_i \, \theta_j] \approx F_{ij}^{-1}.
 \ee

\subsection{Testing the likelihood on simulations}

Although HW test the likelihood reconstruction of the various power spectra 
on both Gaussian and N-body simulations, we have also conducted our own
simulations of the likelihood reconstruction. We do this because of the 
small size of our data fields ($30 \times 30$ arcmin) relative to HW's 
simulated fields, which are about 50 times larger in area. 

\begin{figure}
\centering
\begin{picture}(200,230)
\includegraphics{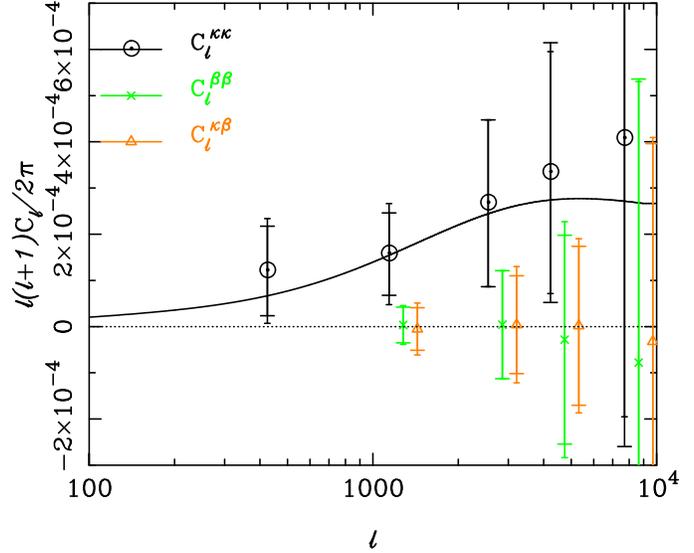}
\end{picture}
\vspace{2mm}
\caption{The recovered band powers of the three spectra, 
$C_{\ell}^{\kappa\kappa}, C_{\ell}^{\beta\beta}$ and $C_{\ell}^{\kappa\beta}$ 
as averaged over 100 Gaussian realisations. Error bars are shown as 
calculated from the run-to-run variations (smaller terminal ends) and 
also as estimated from the Fisher matrix (larger terminal ends). The solid 
curve is the input $\Lambda$CDM power spectrum used to create the simulated 
shear distributions. The $C_\ell^{\beta\beta}$ and $C_\ell^{\kappa\beta}$ points 
have been slightly offset horizontally for clarity.}
\label{sims_cl}
\end{figure}

We have applied the maximum likelihood estimator to one hundred  $30\times30$ 
arcmin fields where the pixelised shear distribution in each field 
is a Gaussian realisation of a shear power spectrum calculated for a 
$\Lambda$CDM cosmology and for all source galaxies lying at $z=1$. 
Following HW's approach, we have added Gaussian distributed noise to 
each of the $20 \times 20$ pixels (pixel size, $1.5 \times 1.5$ arcmin) 
of our shear distribution according to equation (\ref{noisematrix})
where we have taken $\gamma_{\rm rms}=0.4$ for the intrinsic dispersion 
in galaxy ellipticity components. For the number of galaxies in each cell, 
we employ a mean galaxy density of $\overline{n}_g=32.0$ arcmin$^{-2}$, similar 
to the measured galaxy density of our dataset (see Table 1), and we calculate 
the pixel occupation number with $N_{\rm pix}=\overline{n}_g \theta^2_{\rm pix}$
where $\theta_{\rm pix}$ is the pixel side length in arcmin. The results from 
all the simulations are summarised in Fig. {\ref{sims_cl}} where we plot the 
recovered band powers of the three power spectra, 
$C_{\ell}^{\kappa\kappa}, C_{\ell}^{\beta\beta}$ and $C_{\ell}^{\kappa\beta}$, 
averaged over all simulation runs. Here, we plot both the run-to-run errors and 
the errors as estimated from the Fisher matrix which agree well with one another,
supporting the use of the Fisher matrix to estimate errors on the band power
measurements. We see from these simulation results that the maximum likelihood 
reconstruction recovers the input power very accurately for Fourier modes in the range 
$1000 \la \ell \la 5000$. There is some indication from the simulations, 
however, that the method may slightly over-estimate the power for $\ell-$modes 
outside this range. However, this discrepancy is much smaller that the precision
to which we measure the signal. Finally, we note that the likelihood reconstruction 
estimates of the $\beta$-mode power spectrum, $C_{\ell}^{\beta\beta}$ and the 
cross-correlation, $C_{\ell}^{\kappa\beta}$ are both consistent with zero on all 
scales, as expected for the input Gaussian shear model used to create the shear 
distributions.       

\subsection{Maximum likelihood results}

\subsubsection{Shear power spectra}

Having tested the method on simulations, we now apply the maximum likelihood 
reconstruction to our five data fields, CDFS, SGP, FDF, S11 and A901/2. 
As these fields are widely separated on the sky we can treat them as 
independent and optimally combine them afterwards.

\begin{figure*}
\vspace{4.0mm}
\centering
\begin{picture}(220,220)
\includegraphics{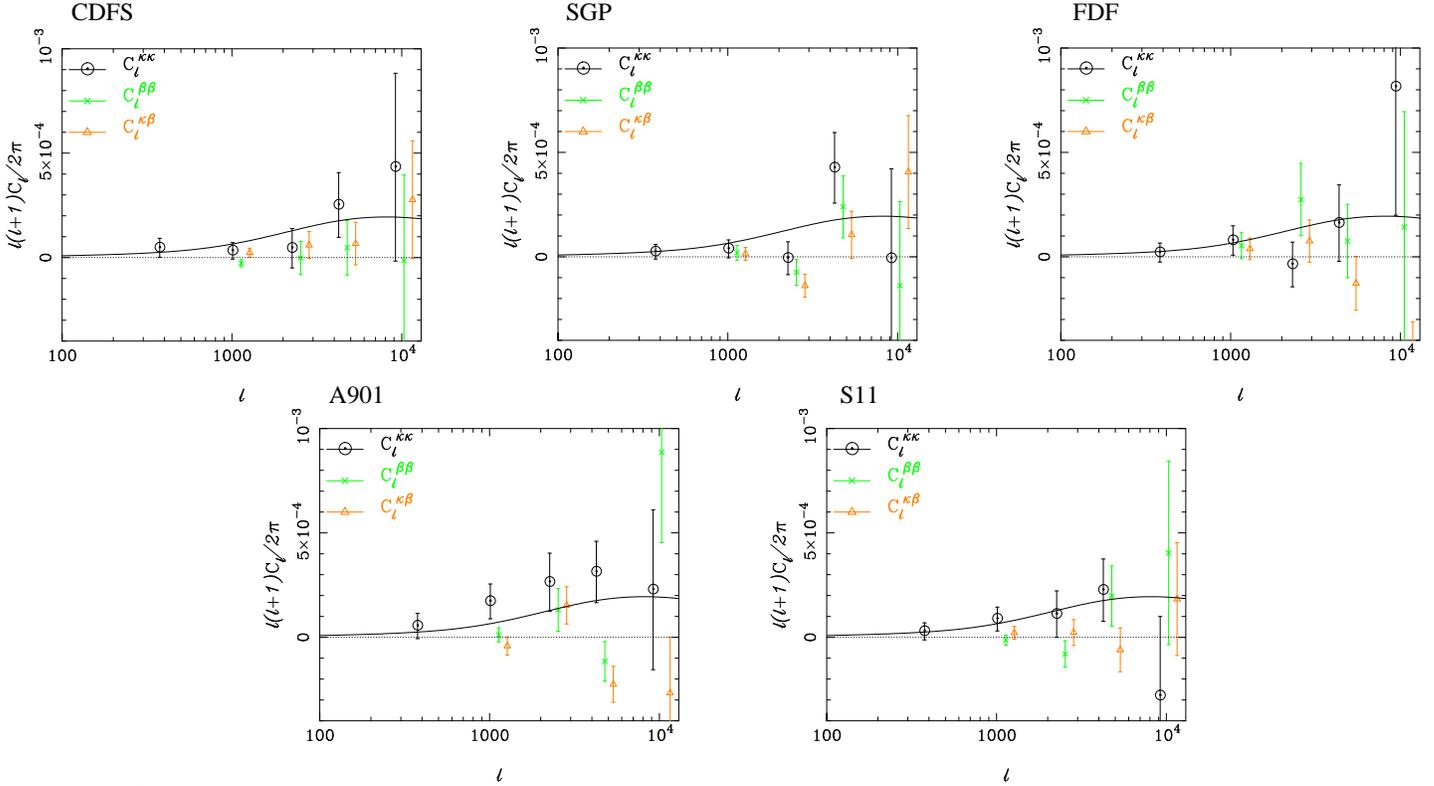}
\end{picture}
\vspace{20mm}
\caption{The cosmic shear power spectra  estimated from each of
the five individual fields in the COMBO-17 dataset. The spectra are for 
(clockwise, from top left) CDFS, SGP, FDF, S11 and A901. Note the much higher 
power recovered from the A901 supercluster field. The $C_\ell^{\beta\beta}$ and 
$C_\ell^{\kappa\beta}$ points have been slightly offset horizontally for clarity.}
\label{all_fields}
\end{figure*}

Fig. \ref{all_fields} shows the results of estimating
$C^{\kappa\kappa}_\ell$, $C^{\kappa \beta }_\ell$ and
$C^{\beta\beta}_\ell$ for the five fields, in 5 band powers. 
Also plotted is a model curve for the $\Lambda$CDM model with
$\Omega_\Lambda=0.7$, $\Omega_m=0.3$, $h=0.68$, normalised to  
$\sigma_8 =0.8$ (see Section 7). The linear power spectrum has been transformed 
to a nonlinear one using the halofit formulae of Smith et al. 
(2002) and, again, we have included the composite redshift distribution 
shown in Fig. \ref{nofz_total} in this calculation.

In general we find that the fields containing clusters, S11 and
A901/2, both yield higher results than the other fields. However, 
as mentioned already, the results for the S11 field seem consistent 
with the dataset as a whole while those for the  A901/2 field clearly 
are not. That the shape is broadly the same for the A901 field is to be 
expected since this part of the nonlinear power spectrum is dominated 
by massive clusters (Cooray \& Hu 2001). At the median redshift of the 
sample, the peak of the nonlinear convergence power spectrum corresponds 
to around $3 h^{-1}$Mpc, the scale on which we expect collapsed clusters 
to have virialised.

The other fields, CDFS, the SGP and FDF, are fields without large 
structures and have correspondingly lower amplitude spectra. 
Interestingly on the smallest scales, $\ell \approx 4000$, the power is 
roughly the same in all five fields.

\begin{figure}
\centering
\begin{picture}(200,230)
\includegraphics{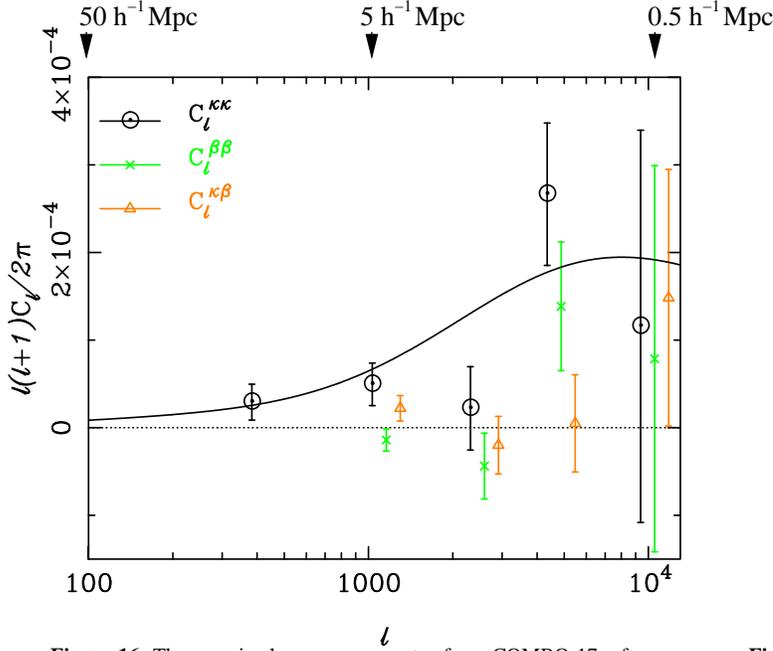}
\end{picture}
\caption{The cosmic shear power spectra from COMBO-17, after excluding the 
A901 supercluster field. Plotted on a linear-log scale are 
$C^{\kappa\kappa}_\ell$ (circles), $C^{\beta \beta}_\ell$ (crosses) 
and $C^{\kappa\beta}_\ell$ (triangles) in 5 band-averaged band powers, 
as a function of multipole, $\ell$, estimated from the optimal combination of a
maximum likelihood analysis of the four COMBO-17 fields, CDFS, SGP,
FDF, and S11. The error bars are estimated from the Fisher
matrix. The solid curve is the shear power spectrum expected for
a $\sigma_8=0.8$ normalised $\Lambda$CDM model. Once again, the 
$C^{\beta \beta}_\ell$ and $C^{\kappa \beta}_\ell$ points have been
slightly horizontally displaced for clarity.} \label{combined_cl}
\end{figure}

Fig. \ref{combined_cl} shows the power from an optimal inverse
weighting combination of the fields excluding A901/2. Details of
the optimally combined power spectra measurements are also given in 
Table \ref{tab3}. 
Three of our five band power measurements agree within their error 
bars with the $\sigma_8=0.8$ normalised $\Lambda$CDM model plotted. 
The high level of power measured at $\ell \sim 4000$ seems to be 
present in each of the fields we have analysed. The largest 
discrepancy between the model and the measurements occurs at 
$\ell=2000$ where less than half the power is found. 

In order to interpret the scales that we are probing with our
power spectrum measurements, we note that, in a flat cosmology, 
the angular size will scale roughly as 
 \be
	R \simeq \frac{2\pi D_A}{\ell},
 \ee
where, for a flat $\Lambda$CDM Universe, the angular diameter 
distance ($D_A$) is well approximated by (Broadhurst, Taylor \& Peacock 1995)
 \be
	D_A(z) \simeq \frac{c}{H_0}\frac{z}{(1+z)(1+3/4\Omega_mz)}.
 \ee
An estimate for the redshift of typical deflectors in the
survey is $z_d \approx 0.5z_m$ and we have estimated $z_m\sim0.85$
(see Section 4.1) for the median redshift of our survey. Combining these 
relations (with $\Omega_m=0.3$) gives the following approximate conversion 
between the physical scale being probed and the Fourier variable on the sky:
 \be
	R \simeq \frac{5130}{\ell} \,\, h^{-1}{\rm Mpc}.
 \ee
This relation gives the approximate physical scale indicated at the 
top of Fig. \ref{combined_cl}.

We can check our results for systematic effects by estimating the
power in the $\beta$-$\beta$ modes, as well as the
$\kappa$-$\beta$ cross correlation. In all of our band powers the
$\beta$-$\beta$ correlation is below the detected signal in $\kappa$ 
and is consistent with zero in all but one band power at $\ell 
\sim 4000$. Similarly the $\kappa$-$\beta$ cross correlation is well
below our measurement of shear power, and is consistent with zero 
except at $\ell \sim 1000$ where a significant detection appears. 
We conclude from the minimal power found in these spectra that our 
results are not strongly contaminated by systematic effects.

\begin{table}
\center
\caption{Details of the maximum likelihood reconstructed band powers as obtained 
from the optimal combination of the four fields, CDFS, S11, SGP and FDF. The first 
two columns list the $\ell$-range of the five band powers. Also listed are the
detections and uncertainties in each band where we quote the band powers, 
$P_{ii}=\ell(\ell+1)C^{ii}_\ell/2\pi$ for $i=\kappa,\beta$. The first band
only has $P_{\kappa\kappa}$ measurements as the $\kappa$ and $\beta$ powers are 
essentially indistinguishable at these large scales.}  
\label{tab3}
\begin{tabular}{@{}lccccc}
\hline
$\ell_{\rm min}$ & $\ell_{\rm max}$ & $P_{\kappa\kappa}(\times10^5)$ & 
$P_{\beta\beta}(\times10^5)$ & $P_{\kappa\beta}(\times10^5)$ \\
\hline
$247$ & $594$  & $2.92\pm2.05$ & -- & -- \\
\\
$595$ & $1790$  & $4.95\pm2.43$ & $-1.45\pm1.26$ & $2.21\pm1.45$ \\
\\
$1791$ & $2986$  & $2.20\pm4.75$ & $-4.38\pm3.76$ & $-1.99\pm3.27$ \\
\\
$2987$ & $6324$ & $26.62\pm8.12$ & $13.85\pm7.34$ & $0.49\pm5.54$ \\
\\
$6325$ & $13906$ & $11.57\pm22.39$ & $7.87\pm22.03$ & $14.83\pm14.64$ \\
\hline
\end{tabular}
\end{table}

\subsubsection{Covariance matrix of band powers}
\begin{figure}
\centering
\begin{picture}(200,230)
\includegraphics{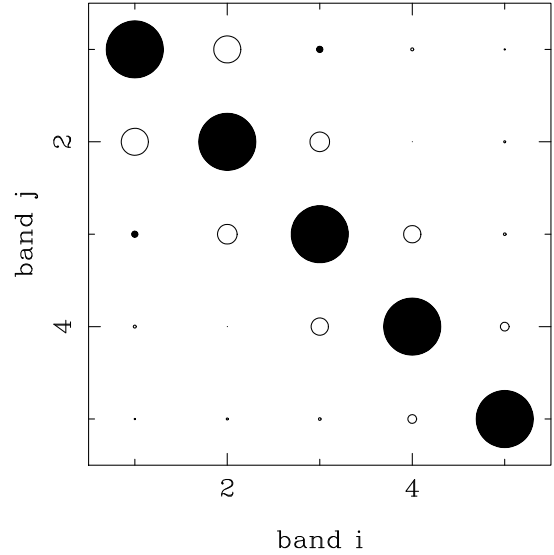}
\end{picture}
\caption{The correlation matrix (see text for details) of the optimally combined 
band power measurements recovered from the maximum likelihood reconstruction 
of $C^{\kappa\kappa}_{\ell}$. The area of each circle is proportional to the degree of 
correlation between bands $i$ and $j$. Filled circles denote that the bands are correlated 
whereas unfilled circles denote an anti-correlation between the bands. The bands are the 
same as those plotted in Fig. \ref{combined_cl}, numbered 1 to 5, in order from left 
to right. Note the small values of the off-diagonal elements, indicating small 
(anti-)correlations between different band power measurements.}
\label{covar_ml}
\end{figure}

In addition to measuring the amplitude of the shear power spectrum, it 
is also important to consider the correlations between band powers. We
can quantify how much the bands are correlated with one another with
the correlation matrix, defined by
 \be
	{\rm Cor}_{ij}=\frac{ {\rm Cov}_{ij} } { \sqrt{ 
        {\rm Cov}_{ii} {\rm Cov}_{jj}}},
\label{corr_matrix}
 \ee
where ${\rm Cov}_{ij}$ is the covariance matrix of the band powers, which 
we measure directly from the data. In Fig. \ref{covar_ml}, we plot the 
correlation matrix of the optimally combined band power measurements shown 
in Fig. \ref{combined_cl}. It is clear from this figure that our band power 
measurements show very little correlation with one another, the biggest effect 
being the slight anti-correlation of neighbouring bands. Thus, our maximum 
likelihood band powers are almost independent measures of power. 

\subsection{Integral power spectra approximations}

As well as a maximum likelihood approach, more direct methods have
been proposed for recovering the shear power spectra. We have
applied one of these direct methods, proposed by Schneider et al.
(2002, hereafter SvWKM), to our data. SvWKM proposed reconstructing
$C^{\kappa\kappa}_{\ell}$ directly from a correlation function
analysis of the data, via an inversion of equation
(\ref{corr2cl}). Pen et al. (2002) have also suggested a similar
approach, but with a somewhat different implementation, which they
apply to the DESCART dataset.

The correlation function estimator, which can also be formulated
in terms of band powers, is not dependent on the spatial
distribution of the shear and, as pointed out by SvWKM
should not suffer from bleeding of power between bands due to
pixelisation of the data. SvWKM test their estimator by 
reconstructing the power spectrum from a fiducial weak lensing survey 
of area, A=25 deg$^2$ from which they assume they have measured the 
correlation functions for $6 \arcsec \le \theta \le 2 \degr$. For 
such a survey, the estimator recovers the input power very accurately 
for $\ell$-modes in the range $200 \la \theta \la 2\times10^5$ and 
SvWKM also measure minimal covariance between their band powers over 
this range. Over the range, $2\pi/\theta_{\rm max}\la\ell\la2\pi/\theta_{\rm min}$,
the correlation function estimator is practically unbiased where 
$\theta_{\rm min}(\theta_{\rm max})$ is the smallest (largest) scale at which the 
correlation functions have been measured. Here we apply a modified version of the 
SvWKM statistic to our dataset. 
If we define two new correlation functions, $\xi_+(\theta)=C_1(\theta)+C_2(\theta)$
and $\xi_-(\theta)=C_1(\theta)-C_2(\theta)$, the correlation function estimator 
of SvWKM can be written as 
 \be
        C_{\ell} =   
        2\pi \int^{\theta_{\rm max}}_{\theta_{\rm min}}  
	\!\!\!\!\! d\theta \, \theta \left[ K_1 \xi_+(\theta) J_0(\ell\theta)  
 	+ K_2 \xi_-(\theta) J_4(\ell\theta) \right], \,\,
	\label{cl_sch}
 \ee
where $J_{0,4}(\ell\theta)$ are the usual Bessel functions and $K_1$ \& 
$K_2$  describe the relative contribution from the $\xi_{+,-}(\theta)$ 
correlators to the integral. In order to achieve an optimal combination of 
$\xi_+$ and $\xi_-$ contributions to the integral, SvWKM 
have constructed a function, $K_1(\ell)$ and imposed the constraint, 
$K_2=1-K_1(\ell)$. However, in order to decompose our signal into $\kappa$ 
and $\beta$ modes, we have chosen $K_1$ and $K_2$ to be constants, 
independent of $\ell$. Setting $K_1=K_2=1/2$ yields a purely $\kappa$-mode 
estimator, whereas if we set $K_1=1/2$ and $K_2=-1/2$, the estimator should 
recover only the $\beta$-mode power. Following SvWKM, we have formulated the 
correlation function estimator in terms of band powers.   
   
\begin{figure}
\centering
\begin{picture}(200,230)
\includegraphics{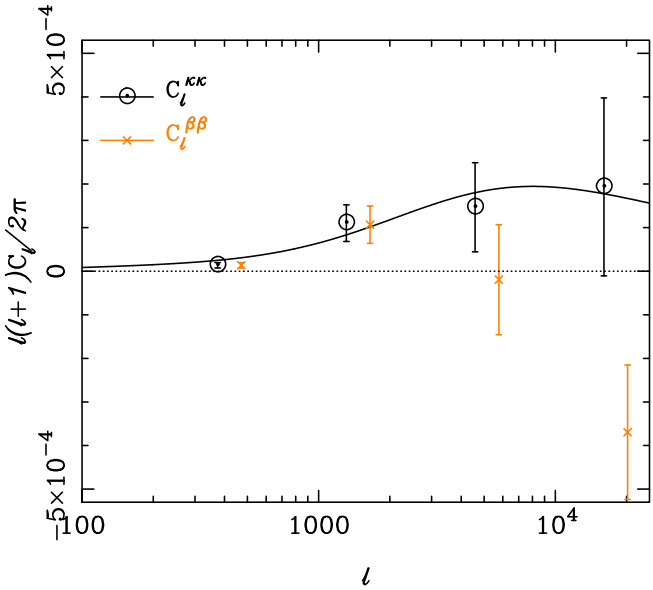}
\end{picture}
\caption{The shear power spectrum, $C^{\kappa\kappa}_{\ell}$ and the 
$\beta$-mode spectrum, $C^{\beta\beta}_{\ell}$ as estimated via a modified 
version (see text for details) of SvWKM's correlation function 
estimator. The $C^{\beta\beta}_{\ell}$ points have been laterally displaced
for clarity. Again, we have excluded the A901 supercluster field for these 
measurements. Also shown is the shear power spectrum expected for a $\sigma_8=0.8$
normalised $\Lambda$CDM model.} \label{combined_schneider}
\end{figure}

Fig. \ref{combined_schneider} shows the result of applying the
correlation function estimator to COMBO-17. Again, we have excluded 
the A901 supercluster field from this calculation. We note that only 
the third band power on this plot is within the stated range of 
validity of the SvWKM estimator. Indeed, this data point is in good 
agreement with the $\Omega_m=0.3$ $\Lambda$CDM model plotted and the 
power in $\beta$ for this band power is consistent with zero. On both 
larger and smaller scales however, as predicted by SvWKM, the estimates 
becomes unreliable, measuring equal $\kappa$ and $\beta$ modes on large 
scales. Ideally, of course, one would measure the correlation functions 
out to as large a scale as possible to extend the range of applicability 
of the estimator. However, for our dataset, we only have correlation 
function measurements in the range $20\arcsec\la\theta\la12\arcmin$ giving us 
a valid $\ell$-range of $1800\la\ell\la5000$. We conclude, therefore, that although 
reconstructing the shear power spectrum from a correlation function analysis can 
be useful for datasets with larger fields, for our purposes, the full maximum 
likelihood analysis is clearly the most suitable approach.  

\section{Cosmological parameter estimation}

\subsection{$\Omega_m$ and $\sigma_8$ from COMBO-17}
Having measured various statistics from the data, we are now in a 
position to use both our measured shear correlation
functions and our reconstructed shear power spectrum estimates 
to obtain a joint measurement of the normalisation of
the mass power spectrum $\sigma_8$, and the matter density
$\Omega_m$. We can achieve this by fitting theoretical shear
correlation functions and power spectra, calculated for particular 
values of these parameters, to our measurements.

Again, we use the fitting functions of Smith et al. (2002) to produce dark
matter power spectra $P_\delta(k,r)$ for values of $\sigma_8$
ranging from 0.1 to 1.5 and $\Omega_m$ from 0.1 to 1.0, exploring all
values on a grid with 0.01 spacing in these parameters. We choose to fix 
$\Omega_m + \Omega_\Lambda = 1$. For all our theoretical power spectra 
and correlation functions, we use $H_0=100h=68 \, {\rm km \, s^{-1} Mpc^{-1}}$ for 
the value of the Hubble constant and $n=1$ for the initial slope of the power 
spectrum of density fluctuations. For each set of parameters, we calculate the 
corresponding shear power spectrum using equation (\ref{eq:gampowspec}). Note 
that we input our composite redshift distribution (Fig. \ref{nofz_total}) into 
the calculation for the shear power spectra. Having calculated the shear power 
spectra we obtain the correlation functions $C_{1,2}$ using equation (\ref{corr2cl}).

We fit the above models to our correlation function and shear power spectrum 
data (where the A901 supercluster field has been excluded in both cases) 
using a $\chi^2$ fitting procedure (c.f. Bacon et al. 2002). We order our
correlation function measurements as a vector 
${\mathbf d}\equiv \{ C_{1}(\theta_{n}), C_{2}(\theta_{n}) \}$, 
where $C_{1,2}(\theta_n)$ is the mean correlation function (averaged over all 
sections) for a given angular separation (see Section 5.1). Similarly, we order 
our theoretical correlation functions as vectors ${\mathbf x}(a)$ with the same 
format, where $a=(\sigma_8,\Omega_m)$.

We use the log-likelihood estimator
\begin{equation}
\chi^{2}=[{\mathbf d}-{\mathbf x}(a)]^{T} {\mathbf
V}^{-1}[{\mathbf d}-{\mathbf x}(a)],
\end{equation}
where
 \be
    {\mathbf V} = \lgl \db \db^t \rgl
 \ee
is the covariance matrix of our correlation function measurements, 
which we measure directly from our data using equation (\ref{covcorr}). 
This estimator is valid in the case of Gaussian errors, which we achieve 
since we have averaged over many sections. We calculate $\chi^2$ for our fine
grid of $(\sigma_8, \Omega_m)$ theoretical correlation functions,
and find the minimum and confidence intervals.

The $\chi^2$ fitting for the shear power spectrum measurements is done in 
exactly the same way as for the correlation function analysis.  

\subsubsection{Correlation function results}
For the correlation function analysis, we find a best fit 
$\sigma_8 (\Omega_m/0.3)^{0.52} = 0.75$, with reduced $\chi^2$ of 1.21. 
The 1$\sigma$ error bar on this value, for a 1 parameter fit, is given 
by the boundary $\Delta \chi^2 = 1.0$, which occurs here at 
$\sigma_8=0.81$ and 0.69. To date, one of the biggest sources of error 
in cosmic shear measurements of cosmological parameters has been the 
uncertainty in the median redshift of the source galaxies. The cosmic 
shear signal scales as $\sigma^2_{\gamma} \propto \sigma_8^{2.5}z^{1.6}$ 
(see e.g. BRE). So the uncertainty in the median redshift contributes to the 
error in $\sigma_8$ as  
\begin{equation}
\left(\frac{\delta\sigma_8'}{\sigma_8'}\right)^2 =
\left(\frac{\delta\sigma_8}{\sigma_8}\right)^2 + 0.64^2
\left(\frac{\delta z}{z}\right)^2.
\label{zerror}
\end{equation}
Using our estimate of the median redshift of the COMBO-17 galaxies, 
$z_m=0.85 \pm 0.05$ (see Section 4.2), we have investigated the extra uncertainty 
introduced into our measurement of $\sigma_8$ by equation (\ref{zerror}).
We find the additional error introduced due to the median redshift 
uncertainty to be 0.01. Thus we obtain a measurement signal-to-noise 
of $\sim9$ for the amplitude of the power spectrum, with a measurement of 
the amplitude 
 \be
	\sigma_8 (\Omega_m/0.3)^{0.52} = 0.75^{+0.08}_{-0.08}.
	\label{constraints_cth}
 \ee
\subsubsection{Shear power spectrum results}
The results for the shear power spectrum analysis yield a slightly lower 
amplitude for the power spectrum normalisation. A good fit to the 
parameter constraints as calculated using our shear power spectrum 
measurements is $\sigma_8 (\Omega_m/0.3)^{0.49} = 0.72$, with a reduced 
$\chi^2$ of 2.21. The additional error due to the uncertainty in $z_s$ is again 
0.01. Our final measurement for the normalisation of the mass power spectrum, as 
calculated from our shear power spectrum measurements is
 \be
	\sigma_8 (\Omega_m/0.3)^{0.49} = 0.72^{+0.08}_{-0.09}.
	\label{constraints_cl}
 \ee 
\vspace{5mm}

\begin{figure}
\centering
\begin{picture}(190,220)
\includegraphics{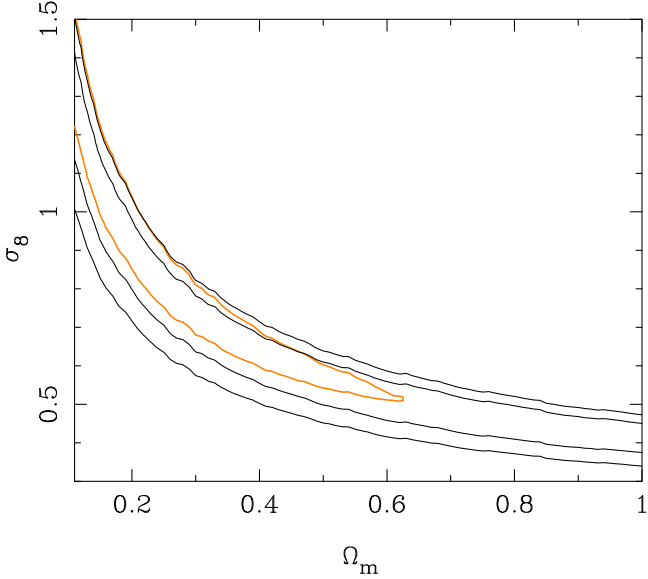}
\end{picture}
\vspace{5mm}
\caption{The likelihood surface of $\sigma_8$ and $\Omega_m$ from COMBO-17
as calculated using our correlation function measurements (lighter contour) and 
as calculated using our shear power spectrum measurements (dark contours). For the 
power spectra measurements we plot the 1 and 2$\sigma$ contours. For the correlation 
functions, we plot only the 1$\sigma$ contour for clarity.} \label{params}
\end{figure}

\noindent
These constraints, as calculated using both of our methods are shown as contours 
in the $\sigma_8 -\Omega_m$ plane in Fig. \ref{params}. Note that in the case of 
the correlation function constraints,  values of $\Omega_m \ga 0.63$ are excluded
at the 1$\sigma$ level.

\subsection{Including the actual redshift distribution}

In order to demonstrate the power of accurate redshift information we have also  
measured parameters from correlation functions which have been calculated using 
only galaxies with reliable redshift estimates. In this case, for our theoretical 
curves, we input the \emph{actual} redshift distribution of the lensed source 
galaxies (i.e. the dotted curve in Fig. \ref{nofz_total}). For the purposes of 
this demonstrative calculation, we have included the A901/2 supercluster field.
We do this simply to increase the number of galaxies -- at present, photometric 
redshifts are only available for the CDFS, S11 and A901 fields and there were too
few galaxies with reliable redshifts in CDFS and S11 alone (19143 galaxies) to 
constrain parameters significantly. Adding the 12506 galaxies with accurate redshifts 
in the A901 field to this sample enabled us to obtain reasonable 
constraints on $\Omega_m$ and $\sigma_8$. The result, therefore, will clearly 
be biased by the A901/2 field and cannot be taken as a measure of the power spectrum 
amplitude. However, it does dramatically demonstrate the increase in accuracy 
attainable with accurate photometric information.  

\begin{figure}
\centering
\begin{picture}(190,220)
\includegraphics{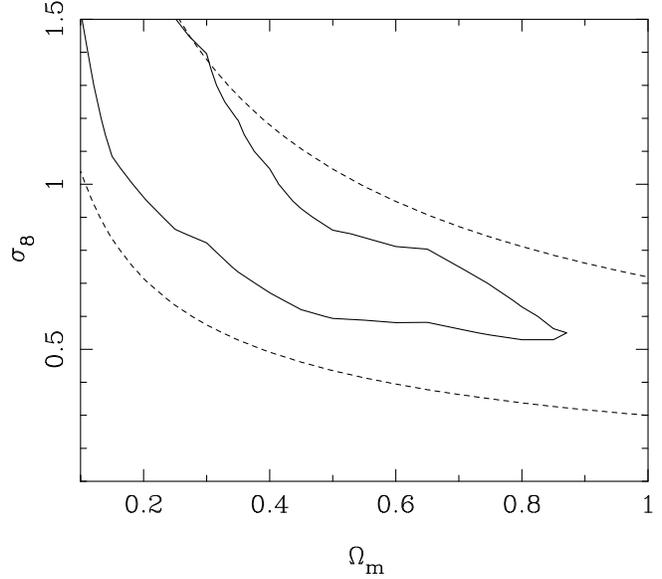}
\end{picture}
\vspace{5mm}
\caption{The likelihood surface of $\sigma_8$ and $\Omega_m$ from
COMBO-17 using correlation functions measured only from galaxies with accurate 
redshift estimates. This subset of galaxies only includes galaxies in the CDFS, 
S11 and A901 fields with {\rm R}-band magnitudes $\le 24.0$.
The 1$\sigma$ contour is plotted as a full line. For comparison, 
we also plot the 1$\sigma$ confidence region (dashed contour) obtained from the 
same correlation functions excluding the redshift information. For this 
calculation the only redshift information we assume is an estimate of the median 
redshift of the survey ($z_m=0.6\pm0.2$).} \label{params_z}
\end{figure}

The results are shown in Fig. \ref{params_z}. For comparison, we also plot 
the constraints obtained from the same set of correlation function measurements 
where we have assumed a median redshift and uncertainty of $z_m=0.6\pm0.2$. 
Here, $z_m=0.6$ is the measured median redshift of galaxies with reliable assigned 
redshifts in the COMBO-17 survey and $\Delta z=0.2$ is typical of the estimated 
uncertainty in the median redshift of cosmic shear surveys to date. In agreement with
the shear power spectrum analysis (see section 6.3), we see the effect of the A901 
supercluster pushing the best-fit $\sigma_8$ value up to $\sim1$ for a matter density
of $\Omega_m=0.3$. Once again, we emphasise that we are not presenting this as a 
measurement of $\sigma_8$ -- we simply wish to demonstrate the dramatic improvement 
in parameter constraints obtainable with the inclusion of accurate photometric 
information. Clearly for accurate parameter estimates, a greater understanding of the 
source redshift distribution must be a priority for cosmic shear surveys in the future.     

\subsection{Combination with the 2dFGRS and CMB experiments}
We can combine the confidence region given by equation (\ref{constraints_cl}) or by
Fig. \ref{params}, with parameter estimations from other sources such as
the 2dF Galaxy Redshift Survey (2dFGRS, Percival et al. 2002) and the various 
CMB experiments (Lewis \& Bridle 2002 \& references therein).

In order to combine our measurements with those from the 2dFGRS and the CMB data,
we first impose some priors on the various datasets. Firstly, we consider 
only flat Universe models ($\Omega_{\Lambda}+\Omega_m=1$) as compelling 
evidence for a flat Universe has already been found from the CMB data 
(e.g. de Bernardis et al. 2002; Pryke et al. 2002; Lewis \& Bridle 2002).  
For the slope of the power spectrum of the primordial density perturbations, 
we take $n_s=1$ and for the baryon fraction $f_b=\Omega_b/\Omega_m$, 
we take $f_b=0.16$. These values are consistent with those measured  
from joint analyses of CMB and Large Scale Structure (LSS) data 
(e.g. Efstathiou et al. 2001; Wang, Tegmark \& Zaldarriaga 2002; Percival et al. 2002; 
Knox, Christensen \& Skordis 2002). 
\begin{figure}
\centering
\begin{picture}(190,220)
\includegraphics{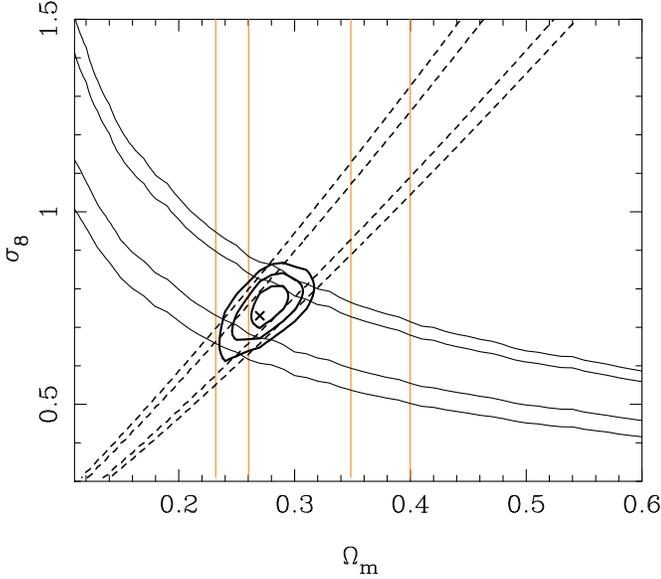}
\end{picture}
\vspace{5mm}
\caption{The likelihood surface of $\sigma_8$ and $\Omega_m$ from
combining the COMBO-17 dataset with the 2dFGRS and the latest CMB constraints. 
The dark thin solid contours are the constraints obtained from our shear power spectrum
analysis described in the previous section. The lighter vertical contours are the 
constraints on $\Omega_m$ obtained from the 2dFGRS where we have applied the priors 
described in the text to the 2dF data. The dashed set of contours
are the constraints from a compilation of six CMB experiments (see text for details) 
where we have assumed an optical depth to reionization of $\tau=0.10$. 
The dark heavy contours are the 1,2 and 3$\sigma$ combined constraints from the 
three methods. The best fit values of $\Omega_m$ and $\sigma_8$ are also indicated.} 
\label{params_2dfc17}
\end{figure}
To display the additional constraints in the $\sigma_8-\Omega_m$ plane, we also need to 
choose a value for the Hubble parameter, $h$. We have taken the  value $h=0.68$, found
from a joint analysis of the CMB and 2dF data for a flat Universe and for a scalar-only 
spectral index, $n_s$ (Percival et al. 2002; Efstathiou et al. 2002). 
Note that this is the same value of $h$ which we adopted when making our model predictions 
in the previous section. Finally, we note that this value is consistent with the HST Key 
Project within the quoted errors (Freedman et al. 2001).     

Having set these various parameters, in the case of the 2dFGRS, for each of our 
grid points in the ($\Omega_m,\sigma_8$) plane, we simply add the 
corresponding likelihood values from Percival et al. (2002) 
to the likelihoods already obtained in the previous section from our shear power 
spectrum data. For the CMB constraints, we have adopted the results of   
Lewis \& Bridle (2002) who provide the following joint constraint on $\Omega_m$ and 
$\sigma_8$ as found from a compilation of six CMB experiments:
 \be
	\nu = \Omega_m h^{2.35} (\sigma_8 e^{-\tau}/0.7)^{-0.84} = 
         0.115 \pm 0.0047.
 \ee 
Here, $\tau$ is the optical depth to reionization. For each point on our 
($\Omega_m,\sigma_8$) grid, we calculate likelihoods for the CMB data as
 \be    
	\chi^2_{\rm CMB} = \frac{ \left( \nu-\Omega_m h^{2.35}
        ( \sigma_8 e^{-\tau}/0.7 )^{-0.84} \right)^2}{ (\delta\nu)^2 }
 \ee
with $\nu=0.115, \delta\nu=0.0047$ and $h=0.68$. Finally, we add the likelihoods from the
various data to get our final measurements of $\Omega_m$ and $\sigma_8$. 
The results of this combination are shown in Figure \ref{params_2dfc17} for an
optical depth of $\tau=0.10$. Note that the 2dFGRS data we have used for this 
estimation constrains $\Omega_m$ only and so the constraints shown on $\sigma_8$ 
come wholly from the cosmic shear and CMB measurements. We measure best-fit values 
of $\Omega_m=0.27^{+0.02}_{-0.01}$ and  $\sigma_8=0.73^{+0.06}_{-0.03}$ from the combined data.

We have also investigated the effect a non-zero $\tau$ has on these measurements. 
We find that increasing/decreasing  $\tau$ has the effect of increasing/decreasing  
the slope of the CMB constraints somewhat, but the resulting combined constraints 
are not altered drastically. For example, for an optical depth of $\tau=0.00$, 
the best-fit $\Omega_m$ and $\sigma_8$ values change to 0.29 and 0.72 respectively, while
increasing $\tau$ to 0.25 changes the best-fit values to $\Omega_m=0.25$ and 
$\sigma_8=0.77$. 

\section{Conclusions}

We have presented an analysis of the cosmic shear signal in the
COMBO-17 survey, a 1.25 square degree survey with excellent data quality,
including photometric redshifts for $\sim 38$\% of our galaxy sample
for three of the five $0.5\degr \times 0.5\degr$ fields.
Our measurements follow a process of careful data reduction,
assessment of the level of telescope-induced shear, and correction for
PSF anisotropy and circularization of galaxies. In this fashion we
reduce the residual systematic effects to a level $<5\%$ of the cosmic
shear signal.

We have measured shear correlation functions and variance-weighted
shear cell-variance for our galaxy sample, detecting the cosmic shear
signal at the $5\sigma$ level. On scales $\ga1$ arcmin, where we are 
confident of the cosmological origin of the signal, we find a 
somewhat lower amplitude than other cosmic shear surveys have measured 
to date.

In contrast to previous cosmic shear surveys, we have substantially reduced
the usual uncertainty introduced into the interpretation of cosmic shear
measurements due to a lack of knowledge of the redshift distribution. 
We have done this by estimating from the data the median 
redshift of the lensed source galaxies to be $z_m=0.85\pm0.05$. 

We have used our catalogues of galaxy positions and shear estimates to
apply a maximum likelihood analysis, obtaining the shear power spectrum
for each COMBO-17 field and for the entire survey. We have measured the
$\kappa$-field power spectrum expected to arise from gravitational shear,
from $\ell=400$ to $10^4$, and have shown that the systematically-induced 
$\beta$-field power spectrum is substantially below our shear signal 
throughout this $\ell$ range. We find that the power spectrum for a
supercluster field has a significantly higher amplitude than that for
random fields, as expected, and we have therefore excluded this field from
our final optimally combined result. Simulations of the maximum likelihood
procedure demonstrate that our method is unbiased to within $\sim10\%$. 
We have measured  the covariance matrix of our shear power spectrum 
band powers directly from the data and find our band powers to be 
essentially independent measured of power.

We have also investigated the use of correlation function estimators 
(e.g. SvWKM) for the shear power spectrum and find, in agreement with 
SvWKM, that their success is dependent on the correlation functions 
being measured over a large range of scales. Such estimators are 
therefore difficult to apply to datasets composed of small fields.  
  
We have used our shear correlation function and power spectrum
measurements to estimate constraints on the cosmological parameters
$\Omega_m$ and $\sigma_8$. We have made $\chi^2$ fits of our measurements
to predicted functions for various values of these cosmological
parameters, and we measure the amplitude of the mass power spectrum,
from our shear power spectrum results to be 
$\sigma_{8} \left( \frac{\Omega_{m}}{0.3} \right)^{0.49} = 0.72 \pm
0.09$ with $0.10<\Omega_m<0.63$, including uncertainties due to
statistical noise, sample variance, and redshift distribution. 

We have demonstrated the potential of including photometric redshift
information by constraining these cosmological parameters with only
those galaxies with assigned redshifts. For this calculation, we have 
input the actual measured redshift distribution into the theoretical 
predictions and have compared the resulting constraints with those 
that would be obtained with only an estimate of the median redshift 
of the survey. We thus demonstrate the dramatic increase in precision 
obtained by including accurate redshift information for the source 
galaxy distribution.   

We have combined our constraints with those from the 2dF Galaxy 
Redshift Survey, and the latest CMB experiments, finding a power 
spectrum normalisation of $\sigma_8=0.73^{+0.06}_{-0.03}$ and a
matter density of $\Omega_m=0.27^{+0.02}_{-0.01}$.

These results (for both the lensing alone, and for the combined constraints) 
are lower than previous constraints found from cosmic shear surveys 
(e.g. Bacon et al. 2002, Refregier et al. 2002, Van Waerbeke et al. 2002, 
Hoekstra et al. 2002) which have, until now found power spectrum normalisations 
of $\sigma_8 \sim 0.85 - 1.0$. However, our measurements are much more in agreement 
with recent cluster abundance estimates of the power spectrum normalisation
(Borgani et al 2001; Seljak 2002; Reiprich \& B\"{o}hringer 2002; Viana, Nichol 
\& Liddle 2002, Allen et al. 2002; Schuecker et al. 2003, Pierpaoli et al. 2002) 
and those found from combining constraints from the 2dFGRS and CMB results 
(e.g. Lahav et al. 2001, Melchiorri \& Silk 2002). Finally, we note that our results 
agree within the $2\sigma$ level with recently announced results from the WMAP satellite 
(Spergel et al. 2003). 

Since this paper was submitted, two other groups have presented cosmic shear 
results yielding somewhat lower values of $\sigma_8$ -- Jarvis et al. 2003 
measure the shear signal from the 75 sq. degree CTIO survey and find a power 
spectrum normalisation of $\sigma_8=0.71$ while Hamana et al. 2002 have used 
the Suprime-Cam instrument on the Subaru telescope to measure the cosmic shear 
signal from 2.1 sq. degs. of deep $R~$band data, with which they constrain the 
power spectrum normalisation to be $\sigma_8=0.69$. The results from both these 
studies are in excellent agreement with the results presented here.      

We have tested our dataset extensively for systematic effects by measuring 
galaxy cross-correlations, star-galaxy correlations and by decomposing the  
cosmic shear signal into its constituent curl and curl-free modes. However,
we have found our dataset to be largely free of any systematic effect that could 
account for the discrepancy found between our parameter constraints and those obtained 
from previous weak lensing studies. One factor, which must be important, is the
increased understanding of the source redshift distribution which we have gained
from the 17-band photometric information contained in the COMBO-17 survey. With 
the increase in area of cosmic shear surveys in the future, this type of 
redshift information is likely to become vital as the uncertainties due to the 
redshift distribution become dominant over other sources of error such as shot 
noise and sampling variance.    

\section*{Acknowledgements}
MLB thanks the University of Edinburgh for a studentship. ANT
thanks the PPARC for an Advanced Fellowship and DJB and MEG are
supported by PPARC Postdoctoral Fellowships. SD thanks the PPARC
for a PDRA grant. CW thanks the PPARC for a PDRA position from the 
rolling grant in observational cosmology at Oxford. We also thank 
Wayne Hu and Martin White for making their original likelihood code 
available, which we adapted to carry out the analysis in Section 6. 
We are grateful to Will Percival for providing us with the 2dFGRS 
data used in Section 7 and we thank Nigel Hambly for providing the 
SuperCOSMOS data for our astrometric calibrations. We thank Peter 
Schneider, Martin White \& Wayne Hu for helpful comments on an earlier 
draft of the paper. Finally, we thank Alan Heavens, Will Percival and 
Ali Higgins for useful discussions.

\section*{References}

 \bib Allen S. W., Fabian A. C., Schmidt R. W., Ebeling H., 2002, 
  MNRAS, accepted (astro-ph 0208394)

 \bib Bacon D. J., Massey R. J., Refregier A., Ellis R., 2002, 
  MNRAS, submitted (astro-ph 0203134)

 \bib Bacon D. J., Refregier A., Ellis R., 2000, MNRAS, 
  318, 625 (BRE)  

 \bib Bacon D. J., Taylor A. N., 2002, MNRAS, submitted (astro-ph/0212266)    

 \bib Barber A., 2002, MNRAS, 335, 909

 \bib Bartelmann M., Narayan, R., 1995, ApJ, 451, 60

 \bib Bartelmann M., Schneider, P., 2001, Physics Reports, 340, 291

 \bib Baugh C. M., Efstathiou G., 1991, MNRAS, 265, 145
     
 \bib Bernardeau F., Van Waerbeke L., Mellier Y., 1997, 
  A\&A, 322, 1  

 \bib Bernstein G., Jarvis M., 2002, AJ, 123, 583

 \bib Bertin E., Arnouts S., 1996, A\&AS, 117, 393
     
 \bib Borgani S. et al., 2001, ApJ, 561, 13

 \bib Broadhurst T. J., Taylor A. N., Peacock J. A., 1995, MNRAS, 299, 895
     
 \bib Brown M. L., Taylor A. N., Hambly N. C., Dye S., 2002, MNRAS, 333, 501
     
 \bib Bunn E., 2002, Phys. Rev. D., 65, 4
     
 \bib Catelan P., Kamionkowski M., Blandford R. D., 2001, MNRAS, 323, 713

 \bib Cohen J. G., Hogg D. W., Blandford R., Cowie L. L., Hu. E., 
  Songaila A., Shopbell P., Richberg K., 2000, ApJ, 538, 29

 \bib Cooray A., Hu W., 2001, ApJ, 554, 56

 \bib Crittenden R., Natarajan P., Pen U., Theuns, T., 
  2001, ApJ, 545, 561
     
 \bib Croft. R. A. C., Metzler C. A., 2001, ApJ, 545, 561 

 \bib de Bernardis P. et al., 2002, ApJ, 564, 559

 \bib Efstathiou G. et al., 2002, MNRAS, 330, L29

 \bib Freedman W. L. et al., 2001, ApJ, 553, 47

 \bib Gray M. E., Taylor A. N., Meisenheimer K., Dye S., 
  Wolf C., Thommes E., 2002, ApJ, 568, 141 (GTM+)
     
 \bib Hamana T. et al., 2002, ApJ, submitted (astro-ph 0210450) 

 \bib Hambly N. et al., 2001, MNRAS, 326, 1279
     
 \bib Heavens A. F., Refregier A., Heymans C., 2000, MNRAS, 319, 649

 \bib Heymans C., Heavens A. F., 2003, MNRAS, 339, 711 

 \bib Hoekstra H., Franx M., Kuijken K., Squires G., 1998, 
  New Astronomy Review, 42, 137

 \bib Hoekstra H., Yee H. K. C., Gladders M. D., 2001, ApJ, 558, L11
     
 \bib Hoekstra H., Yee H. K. C., Gladders M. D., Barrientos L. F.,
  Hall P. B., Leopoldo I., 2002, ApJ, 572, 55

 \bib Hu W., 2000, Phys. Rev. D., 62, 3007 

 \bib Hu W., Keeton C. R., 2002, Phys. Rev. D., 66, 063506 
     
 \bib Hu W., Tegmark M., 1999, ApJ, 514, L65

 \bib Hu W., White M., 2001, ApJ, 554, 67 (HW)
     
 \bib Jain B., Seljak U., 1997, ApJ, 484, 560

 \bib Jarvis M., Bernstein G. M., Fischer P., Smith D., Jain B., 
  Tyson J. A., Wittmann D., 2003, AJ, 125, 1014  

 \bib Kaiser N., 1998, ApJ, 498, 26

 \bib Kaiser N., Squires G., 1993, ApJ, 404, 441
     
 \bib Kaiser N., Squires G., Broadhurst T. J., 1995, 
  ApJ, 449, 460 (KSB)
     
 \bib Kaiser N., Wilson G. Luppino G., Dahle I., 1999, 
  PASP submitted (astro-ph 9907229)
     
 \bib Kamionkowski M., Babul A., Cress C. M., Refregier A., 
  1998, MNRAS, 301, 1064
     
 \bib King L., Schneider P., 2002, A\&A, 396, 411

 \bib Knox L., Christensen N., Skordis C., 2002, ApJ, 563, L95

 \bib Lahav O. et al., 2002, MNRAS, 333, 961

 \bib Lee J., Pen U., 2002, ApJ, 567, L111

 \bib Lewis A., Bridle S., 2002, Phys. Rev. D., 66, 103511

 \bib Luppino G. A., Kaiser N., 1997, ApJ, 475, 20 
     
 \bib Mackey J., White M., Kamionkowski M., 2002, MNRAS, 332, 788
 
 \bib Melchiorri A., Silk, J., 2002, Phys. Rev. D., 66, 041301    
 
 \bib Mellier Y., 1999, ARA\&A, 37, 127

 \bib Pen U., Lee J., Seljak U., 2000, ApJ, 543, L107

 \bib Pen U., Van Waerbeke L., Mellier Y., 2002, ApJ, 567, 31      

 \bib Percival W. J. et al., 2002, MNRAS, 337, 1068     

 \bib Pierpaoli E., Borgani S., Scott D., White M., 2002, MNRAS, 
  submitted (astro-ph 0210567) 

 \bib Pryke C., Halverson W., Leitch E. M., Kovac J., Carlstrom J. E.,
  Holzapfel W. L., Dragovan M., 2002, ApJ, 568, 46

 \bib Reiprich, T. H., B\"{o}hringer H., 2002, ApJ, 567, 716

 \bib Refregier A., Bacon D. J., 2003, MNRAS, 338 48

 \bib Refregier A., Rhodes J., Groth E. J., 2002, ApJ, 572, L131

 \bib Schneider P. Van Waerbeke L., Kilbinger M., Mellier Y., 
  2002, A\&A, 396, 1

 \bib Schuecker P. B\"{o}hringer H., Collins C. A., Guzzo L., 
  2003, A\&A, 398, 867     

 \bib Seljak U., 2002, MNRAS, 337, 769

 \bib Smith R. E., Peacock J. A., Jenkins A., White S. D. M., Frenk C. S.,
  Pearce F. R., Thomas P. A., Efstathiou G., Couchman H. M. P.,
  2002, MNRAS, submitted (astro-ph 0207664) 
     
 \bib Spergel D. N. et al., 2003, ApJ, submitted (astro-ph/0302209)

 \bib Stebbins A., 1997, preprint (astro-ph 9609149)
     
 \bib Taylor A. N., 2001, Phys. Rev. Lett, 
  submitted (astro-ph 0111605)
     
 \bib Taylor A. N., Dye S., Broadhurst T. J., Benitez N.,
  van Kampen E., 1998, ApJ, 501, 539 
     
 \bib Tegmark M., de-Oliviera Costa A., 2002, 
  Phys. Rev. D, 64, 63001 
     
 \bib Tegmark M., Taylor A. N., Heavens A. F., 1997, ApJ, 480, 22 
     
 \bib Van Waerbeke L., Mellier Y., Pell\'{o} R., Pen U., McCracken H. J., 
  Jain B., 2002, A\&A, 393, 369 

 \bib Van Waerbeke L., et al., 2001, A\&A, 374, 757  

 \bib Viana P. T., Nichol R. C., Liddle A. R., 2002, ApJ, 569, L75

 \bib Wang X., Tegmark M., Zaldarriaga M., 2002, 
  Phys. Rev. D., 65, 123001

 \bib Wolf C., Dye S., Kleinheinrich M., Rix H.-W., Meisenheimer K.,
  Wisotzki L., 2001, A\&A, 377, 442   

 \bib Wolf C., Meisenheimer K., Rix H.-W., Borch A., Dye S., 
  Kleinheinrich M., 2003, A\&A, 401, 73    

\label{lastpage}
\end{document}